\documentclass[12pt,a4paper,english]{article}
  \pdfoutput=1
\usepackage{lmodern}
\usepackage[]{hyperref}

\usepackage{ulem}
\usepackage[utf8]{inputenc}
\usepackage{geometry}
\usepackage{color}
\usepackage{float}
\usepackage{amsmath,bm}
\usepackage{amssymb}
\usepackage{graphicx}
\usepackage{setspace}
\usepackage{subfigure}
\usepackage{amsthm}
\usepackage{amsfonts}
\usepackage{braket}
\usepackage{multirow}
\usepackage{cite}
\usepackage{ulem}

\definecolor{purple}{RGB}{204,0,204}

\def\kc#1{\left(#1\right)}
\def\kd#1{\left[#1\right]}
\def\ke#1{\left\{#1\right\}}

\def\be{\begin{equation}}       \def\ee{\end{equation}}
\def\bea{\begin{eqnarray}}      \def\eea{\end{eqnarray}}
\def\ba{\begin{array}}
    \def\ea{\end{array}}
\def\bnum{\begin{enumerate} }
    \def\enum{\end{enumerate}}

\def\=>{\Rightarrow}
\def\>{\rightarrow}

\def\eye2{Fathbb{I}}

\def\eff{\mathrm{eff}}

\makeatletter

\usepackage{slashed}
\usepackage{cite}
\date{}

\makeatother

\usepackage{babel}

\begin{document}

\title{Island in Charged Black Holes}

 \author{Yi Ling$^{1,2}$\thanks{lingy@ihep.ac.cn}, Yuxuan Liu$^{1,2}$\thanks{liuyuxuan@ihep.ac.cn}, Zhuo-Yu Xian$^{3}$\thanks{xianzy@itp.ac.cn}}
\maketitle
\begin{center}
\textsl{$^{1}$Institute of High Energy Physics, Chinese Academy of Sciences, Beijing 100049, China}\\
\textsl{$^{2}$School of Physics, University of Chinese Academy of Sciences,
Beijing 100049, China}
\textsl{$^{3}$Institute of Theoretical Physics, Chinese Academy of Science, Beijing 100190, China}
\par\end{center}
\begin{abstract}
We study the information paradox for the eternal black hole with
charges on a doubly-holographic model in general dimensions, where
the charged black hole on a Planck brane is coupled to the baths
on the conformal boundaries. In the case of weak tension, the
brane can be treated as a probe such that its backreaction to the
bulk is negligible. We analytically calculate the entanglement entropy of the
radiation and obtain the Page curve with the presence of an island
on the brane. For the near-extremal black holes, the growth rate is linear in the temperature. Taking both
Dvali-Gabadadze-Porrati term and nonzero tension into
account, we obtain the numerical solution with backreaction in
four-dimensional spacetime and find the quantum extremal surface
at $t=0$. To guarantee that  a Page curve can be obtained in
general cases, we propose two strategies to impose enough degrees
of freedom on the brane such that the black hole
information paradox can be properly described by the doubly-holographic setup.

\end{abstract}

\newpage
\tableofcontents{}

\section{Introduction}\label{sec_intro}
The black hole information paradox originates from the problem
about whether the information falling into the black hole can eventually sneak out in a unitary fashion. During the evaporation of a black hole, the information carried by the collapsing star appears to
conflict with the nearly thermal spectrum of Hawking radiation by
semi-classical approximation. One possible explanation comes from
quantum information theory. If we assume the
chaotic and unitary
evaporation of a black hole, then the von Neumann entropy of the
radiation may be described by the Page curve, which claims that
the entropy will increase until the Page time and then decrease
 \cite{Page:1993wv,Page:2004xp,Page:2013dx}, which is in
contrast to Hawking's earlier calculation in which the entropy
will keep growing until the black hole totally evaporates
\cite{hawking1974black}. Later, a further debate is raised about
whether the ingoing Hawking radiation at late times is burned up at the horizon
owing to the ``Monogamy of entanglement'', which is known as AMPS
firewall paradox \cite{Almheiri:2012rt}. In order to avoid the
emergence of firewall, the interior of black hole is suggested to
be part of the radiation through an extra geometric
connection (see ``ER=EPR'' conjecture \cite{Maldacena:2013xja}).

Recently AdS/CFT correspondence brings new breakthrough to the
understanding of the Page curve from semi-classical gravity
and sheds light on how the interior of the black hole projects to the radiation
\cite{Almheiri:2018xdw,Penington:2019npb,Almheiri:2019psf}. In
this context, the black hole in AdS spacetime is coupled to a flat
space, where the latter is considered as a thermal bath.

Motivated by the Ryu-Takayanagi formula and its generalization
\cite{Ryu:2006bv,Lewkowycz:2013nqa}, the
fine-grained entropy of a system is calculated by the quantum
extremal surface (QES) \cite{Engelhardt:2014gca}. When applying this proposal to the evaporation of black holes, an island should
appear in the gravity region such that the fine-grained entropy
of the radiation is determined by the island formula \cite{Almheiri:2019hni}
\begin{equation}\label{eq_QESinRad}
S_{\mathcal{R}}=\min_{\mathcal I} \ke{ \mathop{\text{ext}}\limits_{\mathcal I} \kd{ S_{\text{eff}}[\mathcal{R} \cup \mathcal{I}]+\frac{\operatorname{Area}[\partial \mathcal{I}]}{4 G_{N}}}},
\end{equation}
where $\mathcal{R}$ represents the radiation, while the
first term is the entanglement entropy of the region $\mathcal{R} \cup
\mathcal{I}$, and the second term is the geometrical contribution from
the classical gravity. The von Neumann entropy $S_{\mathcal{R}}$ is obtained by the
standard process: extremizing over all possible islands
$\mathcal{I}$, and then taking the minimum of all extremal values.

A doubly-holographic model has been considered in
\cite{Almheiri:2019hni,Chen:2019uhq} (see also
\cite{Chen:2019iro,Balasubramanian:2020hfs,Hashimoto:2020cas,Krishnan:2020fer,Almheiri:2020cfm,Alishahiha:2020qza,Gautason:2020tmk}),
in which the matter field in $2D$ black hole geometry is a
holographic CFT which enjoys the $AdS_3/CFT_2$ correspondence. By
virtue of this, $S_{\text{eff}}$ in (\ref{eq_QESinRad}) is
calculated according to the ordinary HRT formula in $AdS_3/CFT_2$.
At early times, the solution contains no island and the
linear increase of the entropy is contributed from the first
term in (\ref{eq_QESinRad}) as the accumulation of the Hawking
particle pairs. But later, the QES undergoes a phase transition
with the emergence of the island $\mathcal I$. Meanwhile, owing
to the shrinking of the black hole, the decrease of the entropy at
late times is dominated by the second term in
(\ref{eq_QESinRad}). As a result, the whole process is described
by the Page curve and the transition time is the Page time.
Moreover, the conjecture of ER=EPR is realized as the fact that
the island $\mathcal I$ appears in the entanglement wedge of the
radiation \cite{Maldacena:2013xja}.

A similar information paradox occurs when a two-sided black hole
is coupled to two flat baths on each side, and the whole system is in equilibrium \cite{Almheiri:2019yqk}.
Exchanging Hawking modes entangles black holes and baths. However,
according to the subadditivity, the entanglement entropy should be
upper bounded at the Page time $t=t_P$, namely
\begin{align}\label{eq_SubAdd}
S_{\mathcal R} = S_{\mathcal B} \leq S_{\mathcal B_L} + S_{\mathcal B_R},
\end{align}
where $\mathcal B=\mathcal B_L\cup\mathcal B_R$ denotes the
two-sided black hole, with $\mathcal B_L$ and $\mathcal B_R$ being
the black hole on each side, respectively. The equality in
(\ref{eq_SubAdd}) comes from the fact that the whole system
$\mathcal R\cup\mathcal B$ is a pure state. The authors in
\cite{Almheiri:2019yqk} considered the $2D$ eternal black
hole-bath system when the whole holographic system is dual to
Hartle-Hawking state. Its island extends outside the horizon and
the degrees of freedom (d.o.f.) on the island are encoded in the
radiation by a geometric connection
\cite{Almheiri:2019yqk}. A similar conclusion was made
independently by only considering the d.o.f. of the radiation
\cite{Penington:2019kki,Almheiri:2019qdq}.

In \cite{Almheiri:2019psy} (see
\cite{Geng:2020qvw}
for analytically calculable models, \cite{Chen:2020uac,Chen:2020hmv,Hernandez:2020nem} for islands in braneworld and \cite{Akal:2020wfl,Miao:2020oey} for braneworld holography), a nontrivial setup of
the doubly-holographic model for entanglement islands in higher dimensions was established,
where the lower dimensional gravity is replaced by a Planck brane
with Neumann boundary conditions on it
\cite{Takayanagi:2011zk,Chu:2018ntx,Miao:2018qkc}. The solution at
$t=0$ was obtained with the DeTurck trick. Moreover, it was
demonstrated that the islands exist in higher dimensions, and the
main results in
\cite{Penington:2019npb,Almheiri:2019hni,Chen:2019uhq,Almheiri:2019yqk}
can be extended to higher dimensional case as well. However, since
the DeTurck trick is inappropriate for the time-dependent case
\cite{Headrick:2009pv,Dias:2015nua}, in general, the standard Page
curve in higher dimensions is difficult to obtain.

Another issue arises in the context of doubly-holographic setup
when one tries to determine the Page time by the initial entropy
difference between the solution with island and that without island. If the d.o.f. in the
black hole are few compared to the d.o.f. of the baths, then the entropy $S_{\mathcal{R}}$ will
saturate at a fairly low level and the inequality
(\ref{eq_SubAdd}) has to be saturated at $t = 0$ -- see
Sec.~\ref{Prescription} for details. That is to say, the Page time
is $t_P=0$ and thus the Page curve can not be recovered. This phenomenon was firstly noticed  in \cite{Geng:2020qvw} and further elaborated in \cite{Geng:2020fxl}. To avoid
this phenomenon, the essential condition is to input enough d.o.f.
into the black hole at the initial time, which is equivalent to
increasing the entropy in (\ref{eq_QESinRad}). One immediate
resolution is moving the endpoint of the HRT surface away from the
brane \cite{Almheiri:2019psy, Geng:2020qvw}, which is equivalent
to transferring d.o.f. from baths into black holes -- see
Sec.~\ref{sec_RN} for details.

In this paper, we intend to investigate the information paradox in
higher dimensional charged black holes. First of all, it is
crucial to address the information paradox in four or higher
dimensional spacetime. However, to avoid the numerical
difficulties, most of early work on the island paradigm were done
in 2d or 3d gravity theories. \cite{Almheiri:2019psy} made a
significant step to increase the dimensionality by numerical
construction and provided an affirmative answer to the question
whether the information paradox could be averted by the emergence
of an island in higher dimensions, where the AdS-Schwarzschild
black hole is considered as the background specifically. In this
paper we intend to extend this construction to charged black holes
and answer the question whether the island paradigm is a general
resolution to the information paradox for higher dimensional black
holes. It is reasonable to expect that in this context the island
scenario should also be appropriate to describe the entanglement
between the black holes and the baths. More importantly, the
Page curve is absent in \cite{Almheiri:2019psy} though the
relevant discussion and argument support this would be expectable.
Therefore in this paper we intend to specifically obtain the Page
curve in higher dimensions in the weak tension limit where the
backreaction of the brane to the bulk is negligible. We will
analytically calculate the entropy of the radiation and evaluate
the Page time, which could be viewed as a substantial improvement
of the previous work on island in higher
dimensions\cite{Almheiri:2019psy}. In addition, 
taking the charged black hole into account, we are allowed to investigate the Page curve and the Page time at different Hawking temperatures on the unit of the chemical potential, especially in the near extremal case.

The second motivation of
our paper is to propose new strategies to avoid the saturation of
the entropy at the beginning. Rather than transferring d.o.f. from
baths to black holes, we argue that the Planck brane may acquire
enough d.o.f. by increasing the tension and adding a Dvali-Gabadadze-Porrati (DGP) term on
the brane, which leads to the decrease of the Newton constant
$G_N$ \cite{Akal:2020wfl,Miao:2020oey,Chen:2020uac,Chen:2020hmv}
and thus, increasing the entropy in (\ref{eq_QESinRad}). 
A more precise description is from the boundary perspective \cite{Almheiri:2019hni}, where the $d$-dimensional brane theory is dual to the $(d-1)$-dimensional conformal defect. The d.o.f. on the brane and baths are proportional to central charges in the doubly-holographic model \cite{Chen:2020uac}
\begin{align}\label{eq_Ratios}
\frac{c_b}{c}\sim \kc{\frac{l_\eff}{L}}^{d-2}(1+\lambda_b),
\end{align}
where $c_b$ is the central charge of the $(d-1)$-dimensional conformal defect, $c$ is the central charge of the $d$-dimensional bath CFT, $L$ is the AdS radius in the bulk, $l_\eff$ is the effective AdS radius of the induced metric on the brane and $\lambda_b$ is the relative strength of the DGP term. This formula still works near the asymptotic boundary in the presence of black holes. It was shown in \cite{Chen:2020uac} that increasing the tension and adding the DGP term (with positive $\lambda_b$) enlarges the ratio (\ref{eq_Ratios}).
Furthermore, from numerical calculations, we find that the entropy would not be
saturated at the initial time with appropriate tension or DGP term and without transferring any part of the baths into black holes by moving the endpoint of HRT surface away from the brane.

\begin{figure}
  \centering
 \subfigure[]{\label{fig_rn3d}
  \includegraphics[width=195pt]{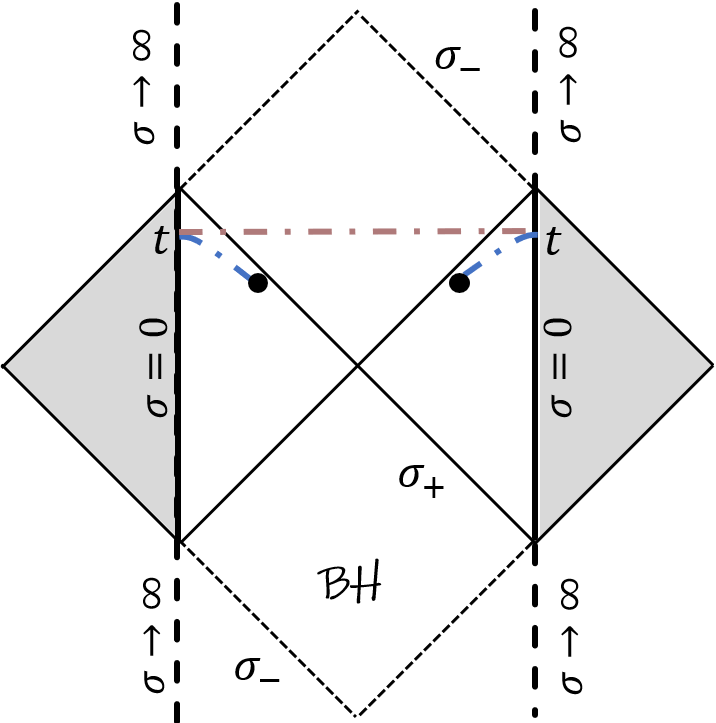}}
  \hspace{0pt}
  \subfigure[]{\label{fig_gr_hb2}
  \includegraphics[height=185pt]{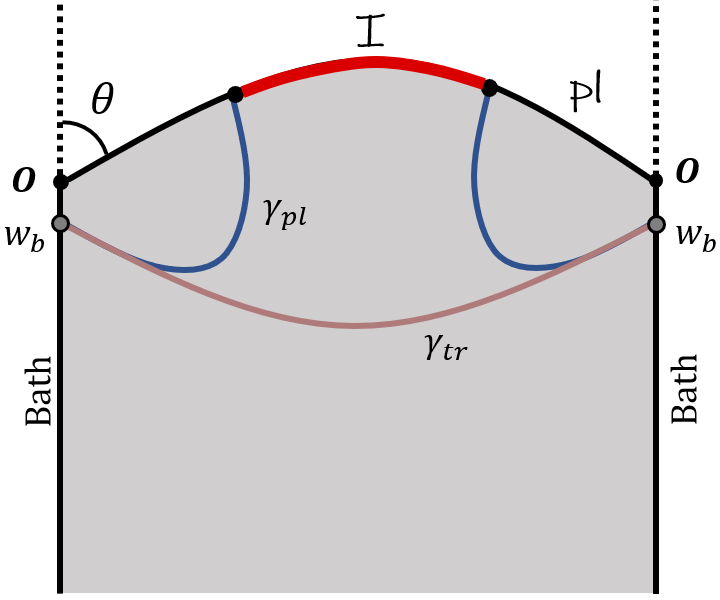}}
\caption{(a): The $d$-dimensional charged black hole is in
equilibrium with two flat baths (colored in gray). The conformal
boundary is located at $\sigma=0$, while the black hole and baths
are distributed in the region with $\sigma>0$ and $\sigma<0$,
respectively. Two candidates of the QES at time $t$ are
 plotted in different colors.
(b):
A sketch of the $(d+1)$-dimensional dual of the $d$-dimensional
holographic system. Here the $d$-dimensional black hole
 is described equivalently by the
Planck brane $\bm{pl}$ in the $(d+1)$-dimensional ambient spacetime. The QES is measured
 by an ordinary HRT surface (which is plotted as a curve either in blue or rose gold) in the
$(d+1)$-dimensional spacetime.}\label{fig_doubleholo}
\end{figure}

The paper is organized as follows. In section \ref{sec_setup}, we
 consider the charged matter on the boundary and build up the
doubly-holographic model. In section \ref{sec_pagecurve},
We neglect the backreaction of the brane and obtain the Page
curve in the weak tension limit in general dimensions. The evolution
behavior at different Hawking temperature will be demonstrated.
In section \ref{sec_numeric}, we take both nonzero tension and
DGP term into account, then obtain the numerical solution with
backreaction in four-dimensional space time and find the quantum
extremal surface at $t=0$. Based on this setup we propose new
strategies to avoid the saturation of (\ref{eq_SubAdd}) at $t=0$,
and then explore their effects on the island in the back-reacted
spacetime.  Our conclusions and discussions are given in section
\ref{sec_Conclusion}.

\section{The doubly-holographic setup}\label{sec_setup}

In this section, we will present the general setup for the island
within the charged eternal black hole. We consider a
$d$-dimensional charged eternal black hole in
$AdS_{d}$ coupled to two flat baths on
each side, with the strongly coupled conformal matter living in
the bulk, as shown in Fig.~\ref{fig_rn3d}. On each side, the
black hole corresponds to the region with $\sigma>0$ and the bath
corresponds to the  region with $\sigma<0$. Moreover, at
$\sigma=0$, we glue the conformal boundary of the $AdS_{d}$ and
flat spacetime together and impose the transparent boundary
condition on the matter sector. With a finite chemical potential
$\mu$, the matter and the black holes carry charges.

The above description can be equivalently pushed forward into a
doubly-holographic setup, where the matter sector is dual to a
$(d+1)$-dimensional spacetime and the $d$-dimensional black hole
is described by a Planck brane $\bm{pl}$ in the bulk, as shown in
Fig.~\ref{fig_gr_hb2}. In this paper, we will adopt the
doubly-holographic setup.

Consider the action of the $(d+1)$-dimensional bulk as
\begin{align}\label{eq_Action}
I=&\frac{1}{16\pi G_N^{(d+1)}} \Bigg[ \int d^{d+1}x
\sqrt{-g}\left(R+\frac{d(d-1)}{L^2}\right)+2\int_{\bm{pl}}d^{d}x\sqrt{-h}\left(K-\alpha\right)\nonumber\\
&+2\int_{\bm{\partial}}d^{d}x\sqrt{-h_{\bm{\partial}}}K_{\bm{\partial}}-\int
d^{d+1}x \sqrt{-g}\frac{1}{2}F^2-2\int_{\bm{pl}\cap \bm{\partial}}
d^{d-1}x \sqrt{-\Sigma}\, \theta\Bigg].
\end{align}
Here $K$ is the extrinsic curvature and the parameter $\alpha$ is
proportional to the tension on the brane $\bm{pl}$, which will be
fixed later. $K_{\bm{\partial}}$ is the extrinsic curvature on the
conformal boundary $\bm{\partial}$. The electromagnetic curvature
is $F=\text{d}A$. The last term is the junction term at the
intersection of the brane $\bm{pl}$ and the conformal boundary
$\bm{\partial}$, where $\theta$ is the angle between the brane and
the boundary, while $\Sigma$ is the metric on $\bm{pl}\cap
\bm{\partial}$. Taking the variation of the action, we obtain the
equations of motion as
\begin{align}
  R_{\mu\nu}+\frac{d}{L^2} g_{\mu\nu}&=\left(T_{\mu\nu}-\frac{T}{d-1}g_{\mu\nu}\right), \quad\text{with}\quad T_{\mu\nu}=F_{\mu a}F_{\nu}{}^{a}-\frac{1}{4}F^2g_{\mu\nu}\label{eq_eineq},\\
  \nabla_\mu F^{\mu\nu}&=0,
\end{align}
where $T$ is the trace of the energy-stress tensor $T_{\mu\nu}$.

\subsection{The Planck brane}

\begin{figure}
  \centering
    \includegraphics[width=250pt]{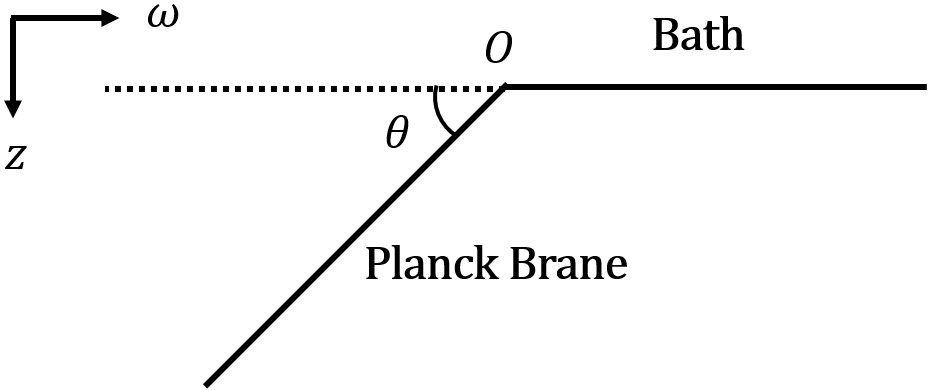}
\caption{A simple setup of Randall-Sundrum
brane\cite{Takayanagi:2011zk}. Here the Planck brane is
anchored on the conformal boundary at $(z,w)=(0,0)$ and penetrates
into the bulk with an angle $\theta$.
  }\label{fig_pb_hb}
  \end{figure}

In AdS/CFT setup with infinite volume, the $(d+1)$-dimensional
bulk is asymptotic to $AdS_{d+1}$ which in Poincaré
coordinates is described by
\begin{equation}
ds^2=\frac{L^2}{z^2}\left(-dt^2+dz^2+dw^2+\sum_{i=1}^{d-2} dw_i^2\right),
\end{equation}
with the conformal boundary at $z=0$.
Let $\theta$ be the angle between the Planck brane and
the conformal boundary as shown in Fig.~\ref{fig_pb_hb}. Then the
Planck brane $\bm{pl}$ is described by the hypersurface
\begin{equation}\label{eq_cons}
 z+w \tan \theta=0,
\end{equation}
near the boundary. One should cut the bulk on the brane
$\bm{pl}$ and restrict it in the region with $z+w \tan
\theta>0$
\cite{Randall:1999vf,Dvali:2000hr,Karch:2000ct,Takayanagi:2011zk}.
We will also impose this constraint (\ref{eq_cons}) deep into
the bulk and find its back-reaction to the geometry.

As for the boundary term in (\ref{eq_Action}), we impose a
Neumann boundary condition on the Planck brane which is
\begin{equation}\label{eq_BCSonBrane}
  K_{ij}-K h_{ij}+\alpha h_{ij}=0,
\end{equation}
where $h_{ij}$ is the induced metric on the brane $\bm{pl}$
. The parameter $\alpha$ in action (\ref{eq_Action}) is
fixed to be a constant by solving (\ref{eq_BCSonBrane})
near the conformal boundary to concrete the tension term on the
brane.

In addition to (\ref{eq_BCSonBrane}), we also impose the Neumann
boundary condition for gauge field $A_\mu$ on the brane
$\bm{pl}$ \cite{Takayanagi:2011zk,Nozaki:2012qd,Chu:2018ntx,Miao:2018qkc}, which is
\begin{align}\label{eq_BCSonBrane1}
  n^\mu F_{\mu \nu} h^\nu{}_i=0,
\end{align}
where $n_\mu$ is the normal vector to the brane and $i$ denotes the coordinates along the brane.

\subsection{The quantum extremal surface}\label{sec_QES}

The von Neumann entropy of the radiation $\mathcal{R}$ in
(\ref{eq_QESinRad}) is measured by the QES . There are two
sorts of candidate for the QES. One sort of candidate is the
disconnected surface, which is represented by the partial Cauchy
surfaces in blue ended with the black dot, while the other is the connected
surface, which is represented by the partial Cauchy surface in
rose gold, as illustrated in Fig.~\ref{fig_rn3d}. Mapping into
the doubly-holographic description, the QES is equivalently
described by a $(d-1)$-dimensional HRT surface
\cite{Almheiri:2019psy}, namely
\begin{align}\label{eq_SR}
S_{\mathcal R}=\min_{\mathcal I} \kd{\frac{\text{Area}(\gamma_{\mathcal I\cup\mathcal R})}{4G_N^{(d+1)}}},
\end{align}
where $\gamma_{\mathcal I\cup\mathcal R}$ is the HRT surface
sharing the boundary with $\mathcal I\cup\mathcal R$, as
illustrated in Fig.~\ref{fig_gr_hb2}. In this figure, each
candidate of the HRT surface corresponds to a co-dimension two
surface $\gamma_{\mathcal{I}\cup\mathcal{R}}$ in the bulk. One is
a trivial surface $\gamma_{tr}$ anchored on the left and right
baths, and the island $\mathcal{I}$ is absent; the other is a
surface $\gamma_{pl}$ anchored on the Planck brane $\bm{pl}$ with
non-trivial island $\mathcal I$ on the brane. The emergence of
island $\mathcal{I}$ keeps the von Neumann entropy
(\ref{eq_QESinRad}) from divergence after the Page time
\cite{Almheiri:2019yqk}.

Consider a HRT surface anchor at $w=w_b$ on the boundary --
Fig.~\ref{fig_gr_hb2}. It measures the entanglement between two
subsystems. One consists of the brane $\bm{pl}$ and part of baths
within the region $0<w\leq w_b$. Conventionally we still call this
subsystem as black hole subsystem hereafter. While the other
subsystem consists of the remaining bath with $w>w_b$, and we
simply call it as the radiation subsystem.

\begin{figure}
  \centering
  \subfigure[]{\label{fig_ProbSol}
  \includegraphics[width=200pt]{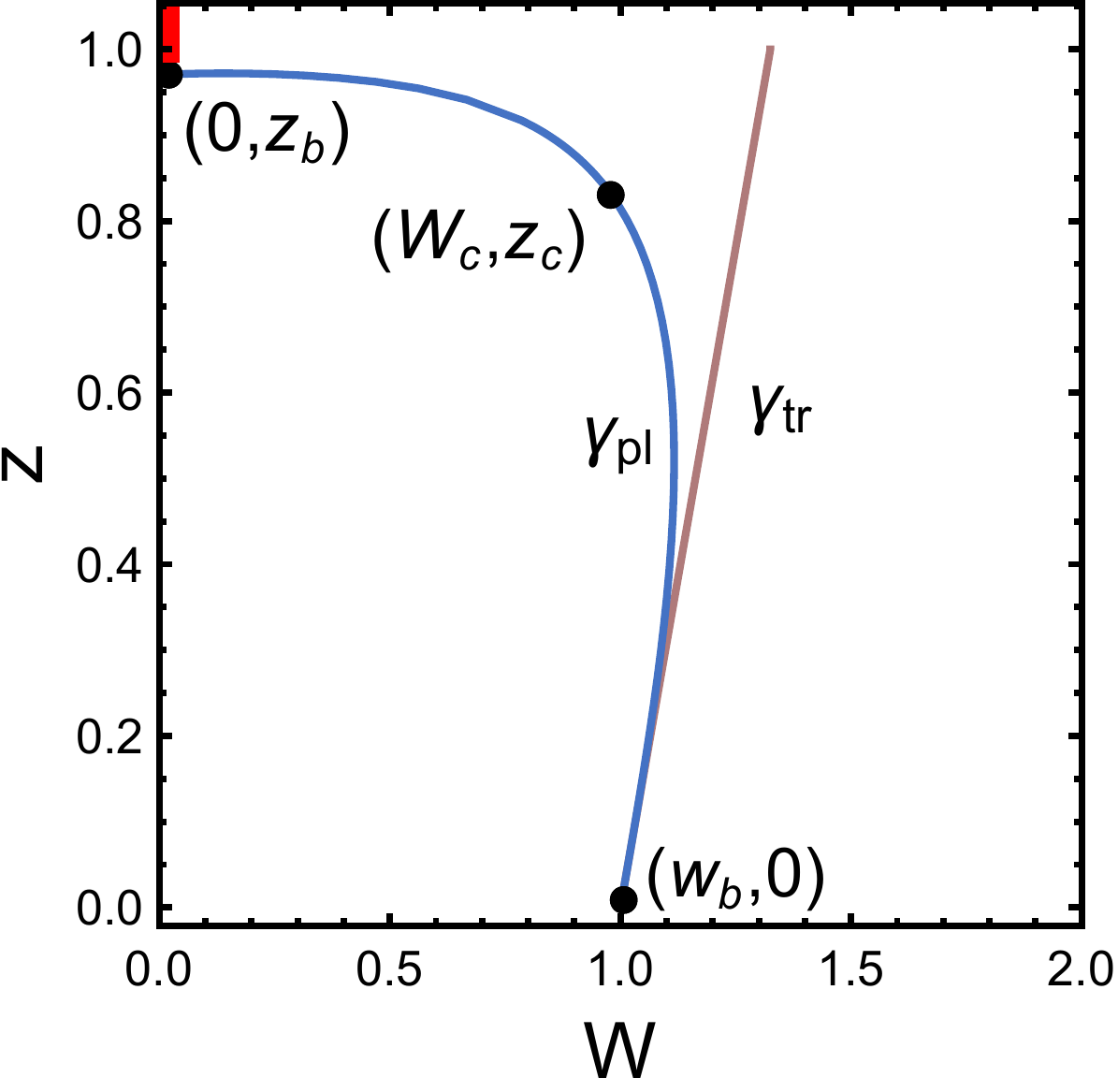}}
  \hspace{20pt}
  \subfigure[]{\label{fig_PenroseDia}
  \includegraphics[width=150pt]{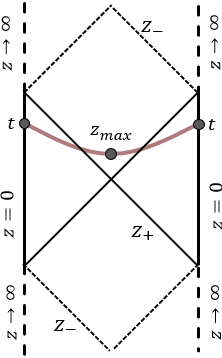}}
\caption{(a): For $\{d,T_h/\mu,\theta,w_b/\mu\}=\{3, 0.45757,2\pi/5, 2\}$, two
candidates $\gamma_{tr}$ and $\gamma_{pl}$ are colored in rose
gold and blue, respectively. Half of the island is colored in red.
(b): The trivial surface $\gamma_{tr}$ is colored in rose gold
at time $t$, where $z_{max}$ is the turning point with
$z'|_{z_{max}}=0$.}
\end{figure}

\section{Entropy density without back-reaction}\label{sec_pagecurve}
Page curve plays a vital role in understanding the information
paradox. In this section, we will firstly compute the entropy
density by figuring out the quantum extremal surface in the weak
tension limit. In such a case, the Planck brane $\bm{pl}$ can be
treated as a probe and its back-reaction to the background
geometry is ignored. Next we consider the growth of the entropy
density with the time. By properly choosing the endpoint of HRT
surface, we plot the Page curve at different Hawking temperature
and evaluate the Page time.

\subsection{The entropy density in the RN black hole}\label{sec_RN}

Define
\begin{align}
W=z\cot\theta+w
\end{align}
such that the Planck brane $\bm{pl}$ is
located at $W=0$ and we only take part of the manifold with $W\geq
0$ into account. 
The tension on the brane is
\begin{equation}\label{eq_a}
\alpha=(d-1)\cos\theta/L.
\end{equation}
When the tension is weak ($\theta \rightarrow
\pi/2$), the brane $\bm{pl}$ applies no back-reaction to the
background, and the geometry can be regarded as the planar
RN-AdS$_{d+1}$, which is
\begin{align}\label{RNBH}
  ds^2&=\frac{L^2}{z^2}\left[-f(z)dt^2+\frac{dz^2}{f(z)}+(dW-\cot\theta dz)^2+\sum_{i=1}^{d-2}dw_i^2\right],\\
  A=&\mu\kc{1-z^{d-2}}dt, \\
  f(z)=&1-
  \left(1+\frac{d-2}{d-1} \mu^2\right) z^d + \frac{d-2}{d-1}\mu^2  z^{2d-2},
\end{align}
where $\mu$ is the chemical
potential of the system on the boundary. The horizon has been rescaled to $z=1$ and one can recover it by transforming coordinates
\begin{align}\label{eq_dim}
\{t,z,W,w_i\}\to \{t,z,W,w_i\}z_h^{-1},\quad i=1,2,...,d-2
\end{align}
In coordinates (\ref{RNBH}), the Hawking temperature is fixed to be
\begin{equation}\label{eq_HawkTemp}
  T_h=\frac{d(d-1)-(d-2)^2\mu^2}{4\pi(d-1)}
\end{equation}
From (\ref{eq_SR}), the entropy density of the radiation subsystem is determined by the minimum
\begin{align}
\tilde S_{\mathcal{R}}=\frac{S_{\mathcal{R}}}{V_{d-2}}=2\min
\kc{\mathop{\text{ext}}\limits_{z_b}\left[\tilde S_{pl}(z_b)
\right], \tilde S_{tr}}.
\end{align}
Here $V_{d-2}=\prod_i^{d-2}\int dw_i$ is the volume of the relevant spatial directions and $z_b$ is the intersection of $\gamma_{pl}$ and the brane. For simplicity, all the above thermodynamic quantities are dimensionless. Following the transformation (\ref{eq_dim}), their dimensions can be recovered as 
\begin{equation}\label{eq_dim2}
\left\{T_h,\mu,V_{d-2},S,\tilde S\right\} \to \left\{T_h \, z_h,\mu \, z_h,V_{d-2}z_h^{2-d},S,\tilde S \,z_h^{d-2}\right\},
\end{equation} 
where $S$ and $\tilde S$ refer to any entropy and its density. Throughout this paper, all the dimensional quantities will be illustrated on the unit of chemical potential $\mu$.

Noticed that
$\gamma_{pl}$ becomes a candidate of the QES only after taking
the extremum. $\mathop{\text{ext}}\limits_{z_b}\left[\tilde
S_{pl}(z_b)\right]$ and $ \tilde S_{tr}$ are the entropy density
contributed by the two candidates respectively -- see
Fig.~\ref{fig_ProbSol}. The entropy density of the radiation subsystem is identified as the minimum of these quantities.

Firstly, we consider the
surface $\gamma_{pl}$ ended on the Planck brane at $z_b$. Two
different parameterizations may be introduced on different
intervals, just as performed in \cite{Almheiri:2019psy}. In
$(W,z)$ plane as illustrated in Fig.~\ref{fig_ProbSol}, we
introduce $W=W(z)$ for the curve segment in $z \in [0,z_c]$, while
for the segment in $W \in [0, W_c]$, we introduce $z=z(W)$
instead, with $z'(W_c)=W'(z_c)^{-1}$.
As a result, the density functional can be evaluated by the area of the surface $\gamma_{pl}$ intersecting the brane at $z_b=z(0)$,
 and the corresponding expression is
\begin{align}
  \tilde S_{pl}(z_b)=&\frac{L^{d-1}}{4 G_N^{(d+1)}} \Bigg(\int_{0}^{z_c} \frac{dz}{z^{d-1}}\sqrt{\frac{f(z) \left[\cot \theta-W'(z)\right]^2+1}{f(z)}}
    \nonumber\\
    &+\int_{0}^{W_c}\frac{dW}{z(W)^{d-1}}\sqrt{\frac{f[z(W)] \left[\cot\theta z'(W)-1\right]^2+z'(W)^2}{f[z(W)]}} \Bigg).
\end{align}
Since $z_b$ depends on $(W_c,z_c)$, $\tilde S_{pl}(z_b)$ can be
view as a function of $z_b$. In Fig.~\ref{fig_ProbSol}, $W=w_b$
is the location where $\gamma_{pl}$ is anchored on the conformal
boundary $z=0$.

\begin{figure}
  \centering
  \subfigure[]{\label{fig_addbt}
    \includegraphics[width=210pt]{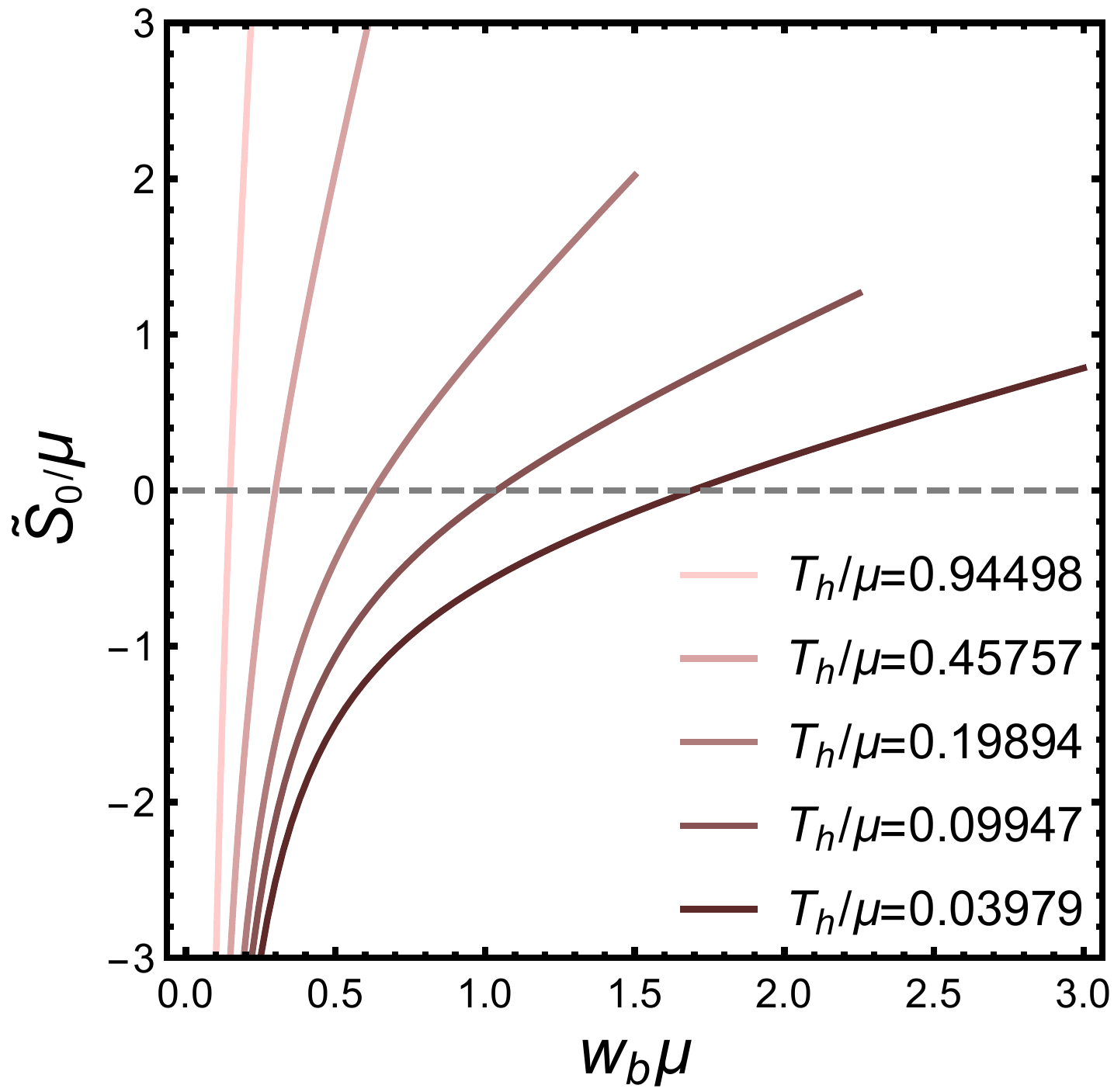}}
    \hspace{0pt}
    \subfigure[]{\label{fig_Probexbyb}
        \includegraphics[width=209pt]{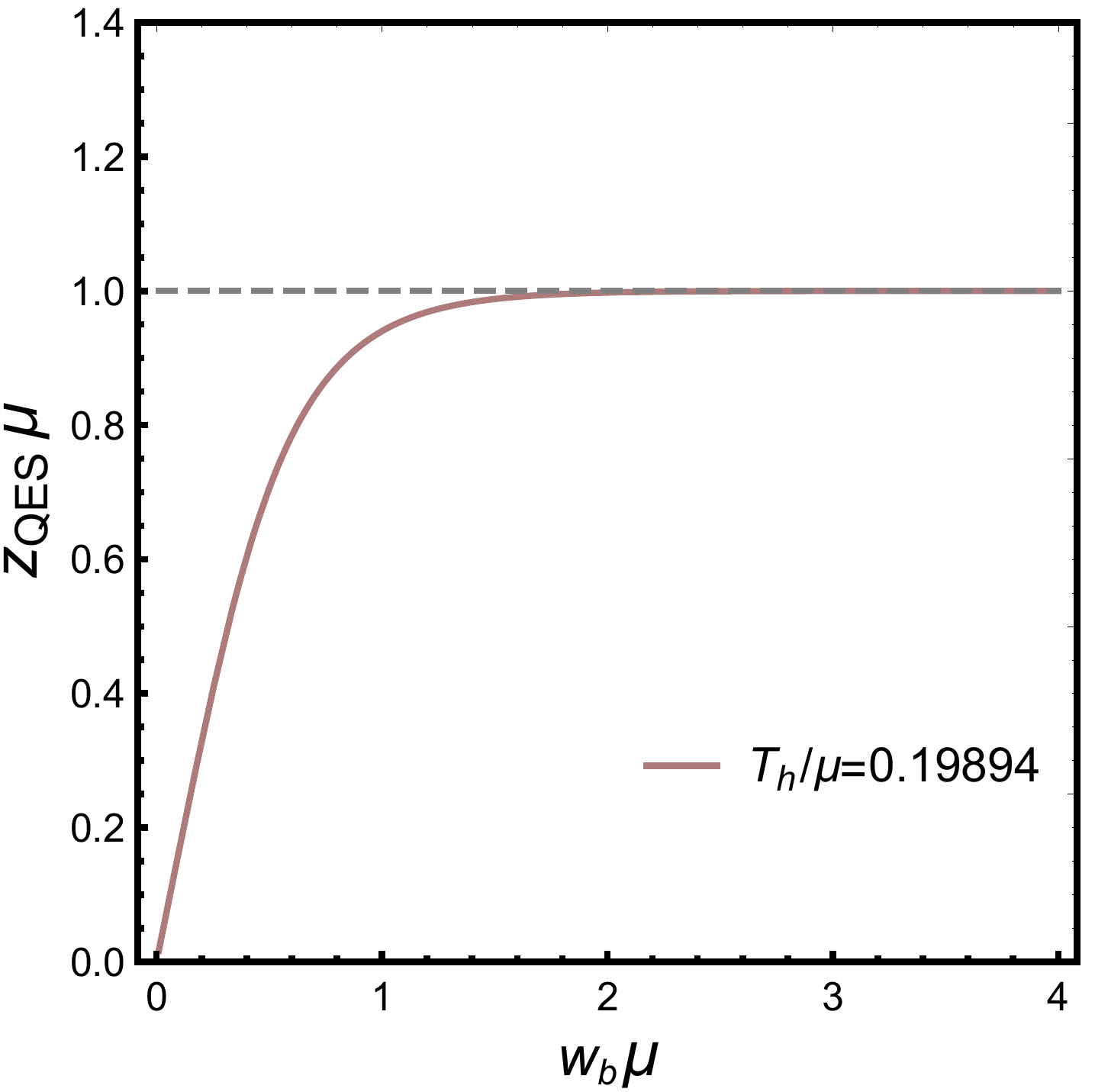}}

    \caption{(a): For $d=3$ and $\theta = \pi/2$, the relations
        between the density difference and the end point of the HRT surface
        are plotted at different temperatures. (b): The location of QES $z_{_{QES}}$ as a function of $w_b \mu$, with
        $\{d,\theta\}=\{3,\pi/2\}$.}
\end{figure}

Next we consider the entropy density associated with the
surface $\gamma_{tr}$ penetrating the horizon as the QES. Since
$\gamma_{tr}$ keeps growing due to the growth of the black hole
interior in the $(d+1)$-dimensional ambient geometry
\cite{Hartman:2013qma}, we express the area functional in the
Eddington-Finkelstein coordinates and the entropy density
$\tilde{S}_{tr}(t)$ becomes
\begin{equation}\label{eq_TBint}
  \tilde S_{tr}(t)=\frac{L^{d-1}}{4G_N^{(d+1)}}\int \frac{ d\xi}{z(\xi)^{d-1}}\sqrt{-v'(\xi ) \left[f(z(\xi )) v'(\xi )+2 z'(\xi)\right]},
\end{equation}
where $\xi$ is the intrinsic parameter on $\gamma_{tr}$
(Fig.~\ref{fig_PenroseDia}) and
$$t=v+\int \frac{dz}{f(z)}.$$

Given the surface $\gamma_{tr}$ which penetrates the horizon at
$t=0$ and the surface $\gamma_{pl}$ which is anchored on the
boundary at any $z_b$, we define their density difference
as
\begin{equation} \label{eq_no}
  \tilde{S}_0[z_b]= 2\left(\tilde{S}_{pl}(z_b) - \tilde{S}_{tr}(0)\right).
\end{equation}
Then the corresponding entropy density difference  between two
QES candidates is given by
\begin{align}\label{eq_SBH}
  \tilde{S}_0 = \mathop{\text{ext}}\limits_{z_b}  \tilde{S}_0[z_b] =2[\tilde S_{pl} (z_{ _{QES}})- \tilde S_{tr}(0)],
\end{align}
where $\tilde{S}_0[z_b]$ exhibits its extremum at $z_b = z_{
_{QES}}$. Then the entropy density of the radiation subsystem
$\tilde{S}_\mathcal{R}$ at the initial time depends on the sign
of $\tilde{S}_0$ as follows: If $\tilde S_0>0$, then
$\gamma_{tr}$ is the HRT surface at $t=0$, which
provides a good starting point for the evolution of the
entanglement between the black hole subsystem and the radiation subsystem. In
this case $\tilde{S}_{tr}(t)$ keeps growing with time $t$ and at
some moment it must reach the entropy density of $\tilde
S_{pl}(z_{ _{QES}})$ \footnote{$\tilde S_{pl}(z_{ _{QES}})$ is
same with the upper bound $S_{\mathcal B_L} + S_{\mathcal B_R}$ in
(\ref{eq_SubAdd}).}. After that, the HRT surface becomes
$\gamma_{pl}$ such that $\tilde S_{\mathcal R}$ stops growing at
the Page time $t_P$. On the contrary, if $\tilde S_0<0$,
then the inequality (\ref{eq_SubAdd}) has to be saturated at
the beginning, indicating that $\gamma_{pl}$ has already been the HRT surface at $t=0$ and no Page curve appears.

In next subsection, we intend to discuss the negativity of
$\tilde{S}_0$ in detail and point out that the negative
$\tilde{S}_0$ results from the lack of d.o.f. in the black
hole subsystem. We will demonstrate that to avoid the negativity
of $\tilde{S}_0$, one may input enough d.o.f. into the black hole
subsystem by choosing relatively large $w_b$, as carried out in
\cite{Geng:2020qvw}.

\subsection{On the difference of the entropy density} \label{sec_TransBath}

First of all, we plot the relation between
 $\tilde S_0$ and the location of the endpoint $w_b$ of
the HRT surface in Fig.~\ref{fig_addbt}. It is noticed that
 generically $\tilde S_0$ exhibits a monotonous behavior with
$w_b$. Specifically it is negative for small endpoint $w_b$, which
leads to the saturation of the inequality (\ref{eq_SubAdd}) at
$t_P=0$. Previously similar phenomenon was observed in
\cite{Geng:2020qvw} with zero tension on the brane. On the other
hand, for large $w_b$ the difference of the entropy density is
always positive and thus appropriate for the Page evolution. Next
we intend to argue that different choices of $w_b$, in effect,
lead to different divisions of the whole system into the black
hole subsystem and the radiation subsystem. Specifically, the
solution with $w_b=0$ measures the entanglement between the brane
and their complementary, namely the whole baths. While for
positive $w_b$, part of baths within $W\leq w_b$ is counted into
the black hole subsystem, and the QES measures the entanglement
between a new black hole subsystem which contains some d.o.f. of
baths and the radiation subsystem at $W>w_b$.

Moreover, the amount of the d.o.f. on the brane encoded in the
radiation subsystem depends on the endpoint $w_b$. To make this
point transparent, we plot the location of QES $z_{ _{QES}}:=z(0)$
as a function of $w_b \mu$ in Fig.~\ref{fig_Probexbyb}.  We find
that the island is always stretching out of the horizon ($z_{
_{QES}}\mu<1$), indicating that besides the interior, the region
near the exterior of the horizon will also be encoded in the
radiation subsystem by the entanglement wedge. This result is
consistent with that in \cite{Almheiri:2019yqk} and reflects the
spirit of the ER$=$EPR proposal, which suggests that two distant
systems are connected by some geometric structure. For $w_b \mu
\rightarrow 0$, the location of QES also approaches $z_{ _{QES}}
\mu \rightarrow 0$, which indicates that the whole region outside
the horizon is likely to be encoded in the radiation subsystem.
While for large $w_b \mu$, $z_{ _{QES}} \mu$ approaches the
horizon, which implies that more d.o.f. in baths are counted into
the black hole subsystem, the less region on the brane is encoded
in the radiation subsystem.

In the remainder of this section, we will choose sufficiently
large $w_b$ to guarantee that $\tilde S_0$ is positive and then
explore the evolution of the von Neumann entropy of the radiation
subsystem. In Sec.~\ref{sec_numeric}, once the backreaction is
taken into account, we will present new strategies to avert the
negativity of $\tilde S_0$, by inputting more d.o.f. on the brane directly, rather than transferring some
d.o.f. from baths.

\subsection{Time evolution of the entropy density}

We are interested in the growth of the entropy density with the
time, thus we define
\begin{equation}\label{eq_DeltaS}
  \Delta \tilde S(t)=\tilde S_{\mathcal{R}}(t)-2 \tilde S_{tr}(0)=
  \begin{cases}
    2\tilde S_{tr}(t)-2\tilde S_{tr}(0), & {t < t_P}\\
    \tilde S_0, & {t \geq t_P}
  \end{cases}
\end{equation}
which is free from the UV divergence and its saturation equals
the density difference $\tilde{S}_0$ as we defined in
Sec.~\ref{sec_RN}.  For large endpoints $w_b$, we have $ \tilde
S_0>0$ and the growth rate of $\Delta  \tilde S(t)$ is obtained
similarly with the method applied in \cite{Carmi:2017jqz}. Since
the integrand of (\ref{eq_TBint}) does not depend explicitly on
$v$, one can derive a conserved quantity as
\begin{equation}\label{eq_C}
  C=\frac{f(z) v'+z'}{z^{d-1}\sqrt{-v' \left[f(z) v'+2 z'\right]}}.
\end{equation}
It is also noticed that the integral shown in (\ref{eq_TBint}) is
invariant under the reparametrization, hence the integrand can
be chosen freely as
\begin{equation}\label{eq_C2}
  \sqrt{-v' \left[f(z) v'+2 z'\right]}=z^{d-1}.
\end{equation}
Substituting (\ref{eq_C}) and (\ref{eq_C2}) into
(\ref{eq_TBint}), we have
\begin{align}
  \frac{d}{dt}\tilde S_{tr}&=\frac{L^{d-1}}{4G_N^{(d+1)}} \frac{\sqrt{-f(z_{max})}}{z_{max}^{d-1}},\\
  t&=\int_0^{z_{max}} dz \frac{C z^{d-1}}{f(z)\sqrt{f(z)+C^2z^{2d-2}}}.
\end{align}
Here $z_{max}$ is the turning point of trivial surface
$\gamma_{tr}$ as shown in Fig.~\ref{fig_PenroseDia}, and the
relation between $z_{max}$ and the conserved quantity $C$ is
given by
\begin{equation}
  f(z_{max})+C^2 z_{max}^{2d-2}=0.
\end{equation}

At later time, the trivial extremal surface $\gamma_{tr}$
tends to surround a special extremal slice $z=z_M$, as shown in
\cite{Hartman:2013qma}. Define
\begin{equation}
  F(z):=\frac{\sqrt{-f(z)}}{z^{d-1}},
\end{equation}
we find that $C^2=F(z_{max})^2$ keeps growing until meeting
the extremum at $z_{max}=z_M$, where we have
  \begin{equation}\label{eq_Fp}
    F'(z_M) = (1-d) z_M^{-d}
    \sqrt{-f\left(z_M\right)}-\frac{z_M^{1-d
    } f'\left(z_M\right)}{2
    \sqrt{-f\left(z_M\right)}} = 0.
  \end{equation}
By solving (\ref{eq_Fp}), we finally obtain the evolution of
the entropy density as
\begin{equation}
  \lim_{t\rightarrow \infty} \frac{d}{dt} \tilde S_{tr} = \frac{L^{d-1}}{4G_N^{(d+1)}}\; F(z_M).
\end{equation}
Next we discuss the growth rate of entropy density for the neutral case
and the charged case separately. Since the saturation occurs at $\tilde S_0$, the Page time is approximately expressed
as
\begin{equation}\label{eq_tpage}
t_P\approx \frac{\tilde S_0}{d\tilde S_{tr}/dt}.
\end{equation}

\begin{itemize}
  \item \textbf{The Neutral Case:}  For $\mu=0$, substituting $f(z)=1-z^d$ into (\ref{eq_Fp}), after recovering the dimension (\ref{eq_dim})(\ref{eq_dim2}),  we have the growth rate of entropy density at late times as
\begin{equation}
  \lim_{t\rightarrow \infty} \frac{d}{dt} \tilde S_{tr} = c_d \frac{L^{d-1}}{4G_N^{(d+1)}} T_h^{d-1}
\end{equation}
where
$c_d=2^{2d+\frac{1}{d}-3}\pi^{d-1}d^{\frac{3}{2}-d}(-1+d)^{\frac{1-d}{d}}(-2+d)^{\frac{d-2}{2d}}$.
For $d=2$, $c_d=2\pi$ and the growth rate is proportional to Hawking
temperature of the black hole, which is exactly in agreement
with the result in \cite{Almheiri:2019yqk,Hartman:2013qma}.

Due to the exchanging of Hawking modes, the entanglement entropy
of the radiation subsystem grows linearly during most of time at a
rate proportional to $T_h^{d-1}$. If there was no island, the
entanglement entropy would keep growing and finally exceeding the
maximal entropy the black hole subsystem allowed to contain. It
would be an information paradox similar to the version of
evaporating black hole. The formation of quantum extremal
surface with island at Page time resolves this paradox, since
some d.o.f. on the brane are encoded in the radiation subsystem
and the entanglement entropy stops growing and becomes
saturated.
\end{itemize}

\begin{figure}
  \centering
  \includegraphics[width=400pt]{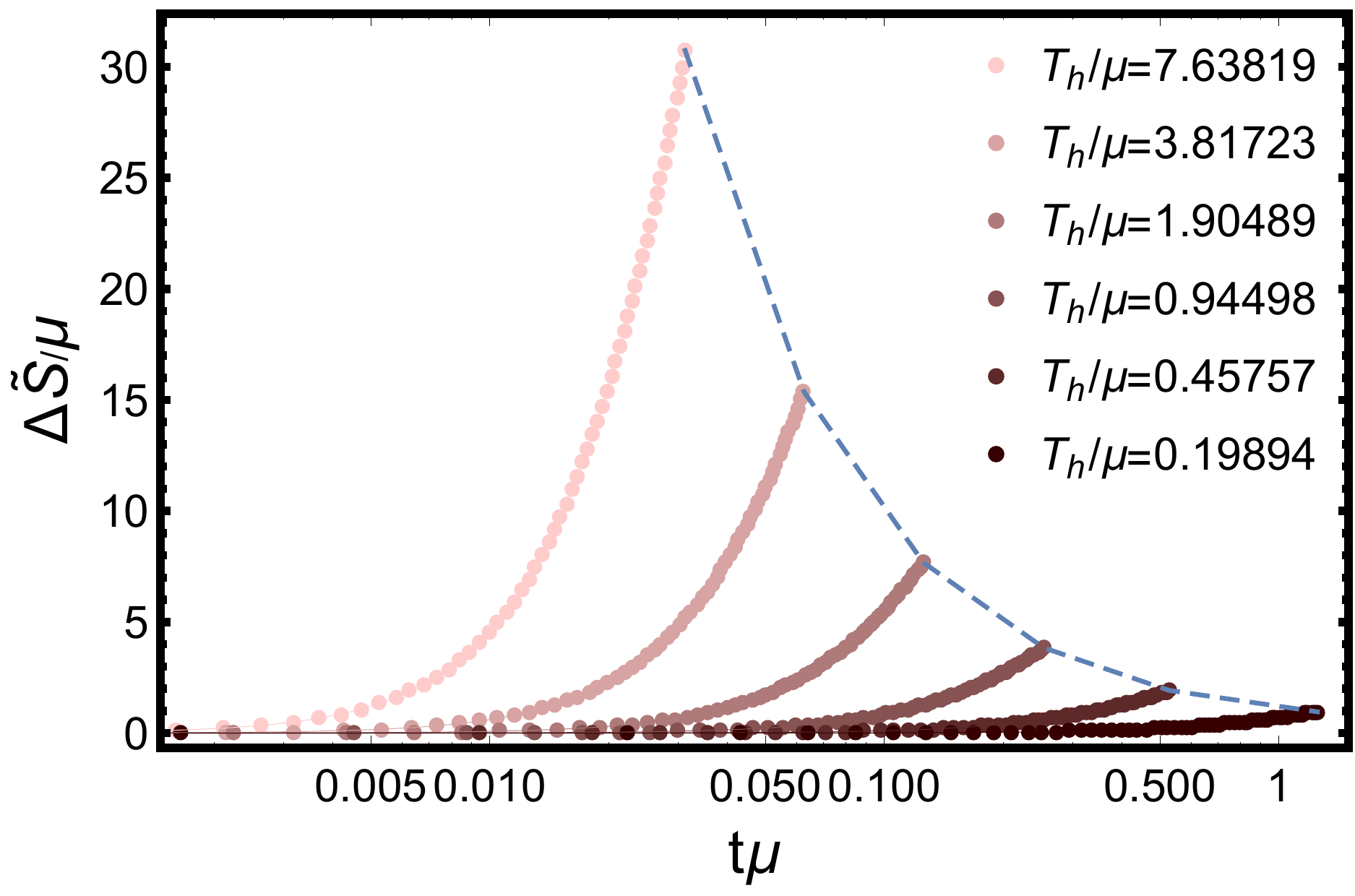}

\caption{For $\{d,w_b,L,\theta\}=\{3,1,1,9\pi/20\}$, the Page
curves are plotted for different Hawking temperatures.
The dashed curve in blue connects the Page time $t=t_P$
of each Page curve. In the plot, The Newton constant is fixed to
be $\frac{L^2}{4G_N^{(4)}}=1$.
}
\label{fig_NewPageCurve}
\end{figure}

\begin{itemize}
  \item \textbf{The Charged Case:} Recall that in general dimensions, when turning on the chemical potential, the blackening factor becomes
  \begin{equation}
    f(z)=1-\left(1+\frac{d-2}{d-1}\mu^2\right)z^d+\frac{d-2}{d-1}\mu^2z^{2d-2}.
  \end{equation}
The growth rate of entanglement entropy with $d\geq3$ at late times
is obtained by substituting the above equation into (\ref{eq_Fp}),
which is
  \begin{align}
    \lim_{t\rightarrow \infty} \frac{d}{dt} \tilde{S}_{tr} = &\frac{L^{d-1}}{4G_N^{(d+1)}}\left(\frac{2(d-1)^2}{(d-2)(d-1+(d-2)\mu^2)}\right)^{\frac{1-d}{d}}\nonumber\\
    & \sqrt{-1+\frac{2(d-1)}{d-2}-\frac{d-2}{d-1}\left(\frac{2(d-1)^2}{(d-2)(d-1+(d-2)\mu^2)}\right)^{\frac{2(d-1)}{d}}\mu^2}.  \end{align}
    Specially, after recovering the dimension (\ref{eq_dim})(\ref{eq_dim2}), the growth rate for the near-extremal black hole is
\begin{align}
  \lim_{t\rightarrow \infty \atop T_h \to 0} \frac{d}{dt} \tilde{S}_{tr} = \frac{L^{d-1}}{4G_N^{(d+1)}} 2\pi \sqrt{\frac{1}{d(d-1)}} \kd{\frac{d(d-1)}{(d-2)^2}}^{\frac{2-d}{2}}T_h \mu^{d-2}.
\end{align}
The linear-$T_h$ behavior comes from the AdS$_2\times R^{d-1}$ space, which is the near horizon geometry of the near-extremal black hole.

Some numerical results are shown in Fig.~\ref{fig_NewPageCurve}. When $T_h/\mu\gg1$, the entropy density approach the neutral case as mentioned above, which has a high growth rate and a large saturated value. When $T_h/\mu\ll1$, the entropy density has a low growth rate and a small saturated value. At the extremal case $T/\mu\to0$, the evolution is nearly frozen and the entanglement entropy barely grows.
\end{itemize}

Since the entanglement between the black hole and the radiation subsystem
is built up by the exchanging of Hawking modes before the Page
time $t_P$, the phenomenon that entropy increases rapidly at
higher temperatures indicates that the higher the Hawking
temperature is, the higher the rate of exchanging is.

\section{Entropy with back-reaction}\label{sec_numeric}

In this section we will consider the QES in the presence of the
island when the back-reaction of the Plank brane to the bulk is
taken into account for $d=3$. First of all, we will introduce a
DGP term on the brane, and discuss its effects on the setup. Next,
we will provide new strategies to enhance the d.o.f on the brane
such that the entropy density difference is guaranteed to be
positive at the initial time. After that, we will apply the
DeTurck trick to handle the static equations of motion and find
the numerical solution to the background with
backreaction at $t=0$ via spectrum method.
Finally, the effects of the tension and the DGP term on the
evolution will be investigated.

\subsection{The DGP term}
We introduce the action of a $d$-dimensional brane into (\ref{eq_Action}) as
\begin{align}\label{eq_Action2}
I_b= \frac{1}{16 \pi G_b^{(d)}}\int d^d x
\sqrt{-h}R_h+\frac{1}{8\pi G_b^{(d)}}\int_{\bm{pl}\cap \bm{\partial}} d^{d-1}x \sqrt{-\Sigma} \, k.
\end{align}
The first term on the r.h.s is the DGP term
\cite{Chen:2020uac}, where $G_b^{(d)}$ is the additional Newton
constant on the brane and $R_h$ is the intrinsic curvature
on the brane. The second term is the junction term at the
intersection $\bm{pl}\cap \bm{\partial}$ of the brane and the conformal boundary, where
 $\Sigma$ is the metric on
$\bm{pl}\cap \bm{\partial}$ and $k$ is the extrinsic curvature on $\bm{pl}\cap \bm{\partial}$.

For the boundary term in (\ref{eq_Action}) and (\ref{eq_Action2}), the new
Neumann boundary conditions are
\begin{equation}\label{eq_BCSonBrane2}
  K_{ij}-K h_{ij}+\alpha h_{ij}=\lambda L \left[\frac{1}{2}R_h h_{ij}-(R_{h}){}_{ij}\right],
\end{equation}
where $\lambda \equiv \frac{G_N^{(d+1)}}{G_b^{(d)}L}$  can be regarded as the effective coupling of the
DGP term. The parameter $\alpha$ in action (\ref{eq_Action}) is
now obtained by solving (\ref{eq_BCSonBrane2})
near the conformal boundary to concrete the tension term on the
brane.

The von Neumann entropy of the radiation subsystem $\mathcal{R}$ in
(\ref{eq_QESinRad}) is now equivalently described by a $(d-1)$-dimensional HRT surface and a
lower-dimensional area term on the brane $\bm{pl}$ \cite{Chen:2020uac,Chen:2020hmv,Hernandez:2020nem}, namely
\begin{align}\label{eq_SR2}
S_{\mathcal R}= \frac{1}{4G_N^{(d+1)}}\, \min_{\mathcal I}\kd{\mathop{\text{ext}}\limits_{\mathcal I}[\textbf{Area}(\gamma_{\mathcal I\cup\mathcal R}) + \lambda L \,\textbf{Area}(\partial\mathcal I)]},
\end{align}
where $\gamma_{\mathcal I\cup\mathcal R}$ is the HRT surface sharing the boundary with $\mathcal I\cup\mathcal R$.

\subsection{The metric ansatz}

We introduce the Deturck method \cite{Dias:2015nua} to
numerically solve the background in the presence of the Planck
brane in this subsection, and the numerical results for the QES
over such backgrounds will be presented in next subsection.
Instead of solving (\ref{eq_eineq}) directly, we solve the
so-called Einstein-DeTurck equation, which is
\begin{align}
  R_{\mu\nu}+3\, g_{\mu\nu}&=\left(T_{\mu\nu}-\frac{T}{2}g_{\mu\nu}\right)+\nabla_{(\mu}\xi_{\nu)},
\end{align}
where $$\xi^{\mu}:=\left[\Gamma_{\nu \sigma}^{\mu}(g)-\Gamma_{\nu
\sigma}^{\mu}(\bar{g})\right] g^{\nu \sigma}$$ is the DeTurck
vector and $\bar{g}$ is the reference metric, which is
required to satisfy the same boundary conditions as $g$ only on
Dirichlet boundaries, but not on Neumann boundaries
\cite{Almheiri:2019psy}.

\begin{table}
  \centering
  \begin{tabular}{|c| c| c |c |c| c |c|}
    \hline
      & 1 & 2 & 3 & 4 & 5 & 6 \\
    \hline
    $\mathbf{y=1}$ & $Q_1=1$ & $Q_2=1$ & $Q_3=\cot\theta$ & $Q_4=1$ & $Q_5=1$ & $\psi=\mu$\\
    \hline
    $\mathbf{y=0}$ & $\partial_y Q_1=0$ & $\partial_y Q_2=0$ & $\partial_y Q_3=0$ & $\partial_y Q_4=0$ & $\partial_y Q_5=0$ & $\partial_y \psi=0$\\
    \hline
    $\mathbf{x=1}$ & $Q_1=1$ & $Q_2=1$ & $Q_3=\cot\theta$ & $Q_4=1$ & $Q_5=1$ & $\psi=\mu$\\
    \hline
    $\mathbf{x=0}$  & $n^\mu F_{\mu \nu} h^\nu{}_i=0$ & $n_\mu \xi^{\mu}=0$ & $Q_3=\cot\theta$ & \multicolumn{3}{c|}{Equation (\ref{eq_BCSonBrane2})}\\
    \hline
  \end{tabular}
  \caption{Boundary conditions.} \label{tab_BCS}
\end{table}

Now we introduce the metric ansatz and the boundary conditions
in the doubly-holographic setup. For $W\rightarrow\infty$, the
ambient geometry is asymptotic to $4D$ planar RN-AdS black
(\ref{RNBH}) with $d=3$. Furthermore, for numerical
convenience we define two new coordinates in the same way as
applied in \cite{Almheiri:2019psy}, which are
\begin{align}
  x=\frac{W}{1+W}\qquad \text{and} \qquad y=\sqrt{1-z}.
\end{align}
The domain of the first coordinate $x$ is compact, while the second coordinate $y$ keeps the metric from divergence at the outer horizon $y_+=0$.

When the back-reaction of the Planck brane is taken into
account, the translational symmetry along $x$ direction
is broken. Therefore, the most general ansatz of the background
is
\begin{align}\label{eq_Background}
  ds^2=&\frac{L^2}{(1-y^2)^2}\left[-y^2 P(y) Q_1 dt^2 +\frac{4 Q_2}{P(y)}dy^2+
  \frac{Q_4}{(1-x)^4}\left(dx+ 2 y (1-x)^2 Q_3 dy \right)^2+Q_5dw_1^2\right]\\
  A=&y^2\,\psi\,dt, \label{eq_GaugeField}\\
  P(y)=&2-y^2+(1-y^2)^2-\frac{1}{2}(1-y^2)^3\mu^2.
\end{align}
where $\ke{Q_1,Q_2,Q_3,Q_4,Q_5,\psi}$ are the functions of
$\kc{x,y}$. All the boundary conditions are listed in
Tab.~\ref{tab_BCS} , with
\begin{equation}\label{eq_a1}
  \alpha=\left(2\cos \theta -\lambda \sin^2 \theta\right) / L.
\end{equation}
Moreover, the boundary conditions at the
horizon $y=0$ also imply that $Q_1(x,0)=Q_2(x,0)$, which
fixes the temperature of the black hole  (\ref{eq_HawkTemp}) as
\begin{equation}
  T_h=\frac{6-\mu^2}{8\pi}.
\end{equation}

\begin{figure}
  \centering
  \includegraphics[width=150pt]{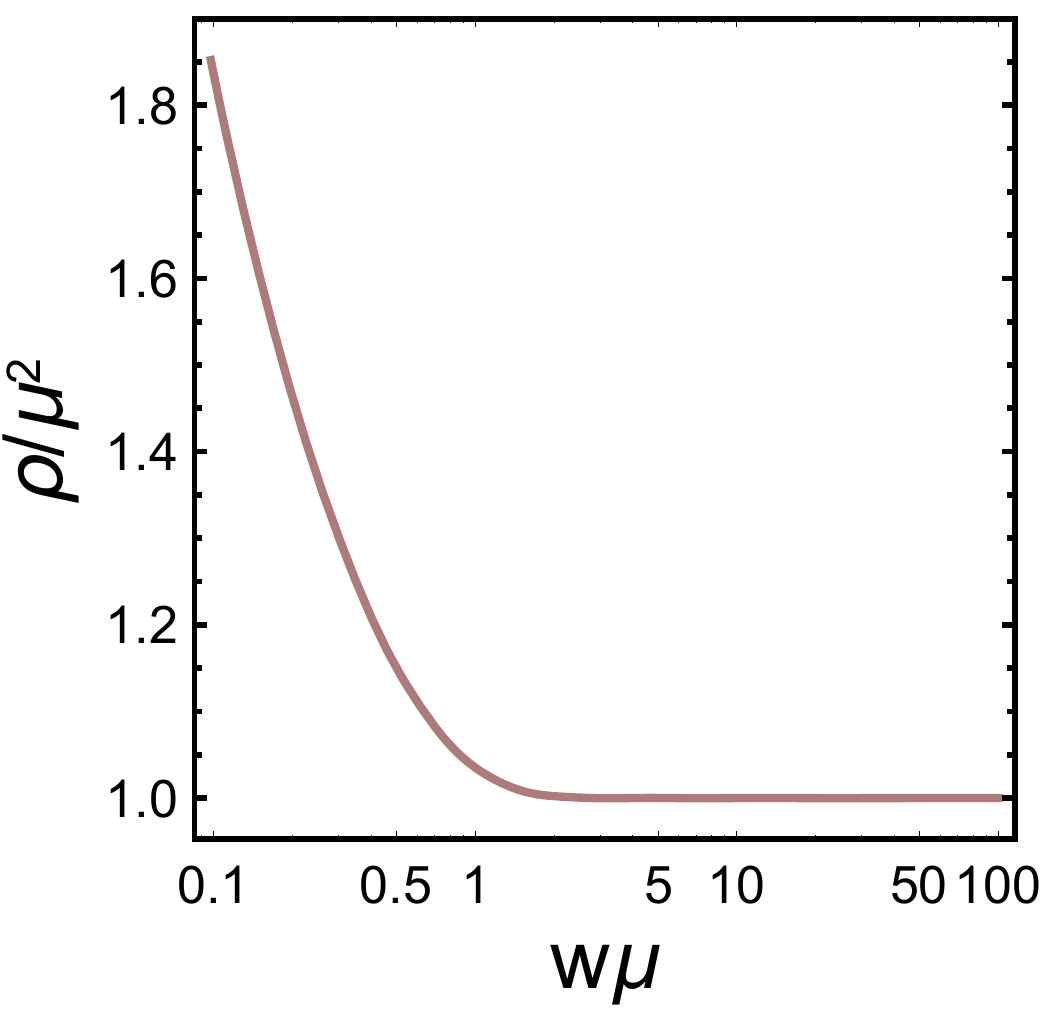}
  \caption{The charge density in the case for $\{T_h/\mu,\theta,\lambda\}=\{0.19894,\pi/4,1\}$.}\label{fig7}
\end{figure}

The reference metric $\bar{g}$ is given by $Q_1=Q_2=Q_4=Q_5=1$ and
$ Q_3=\cot\theta$. In this case, the corresponding charge density
$\rho$ decays
 from the Planck brane $\bm{pl}$ to
the region deep into the bath as shown in Fig.~\ref{fig7}.

\subsection{New prescriptions for the saturation at $t=0$}\label{Prescription}

Firstly, let us elaborate the whole evaporation process in a
specific manner. At the beginning of the evolution, two black
holes $\mathcal B_L$ and $\mathcal B_R$ are entangled with each
other. As time passes by, the black holes interact with
environment (radiation) by exchanging the hawking modes, and the
von Neumann entropy of the radiation increases. At the Page time,
two black holes are disentangled with each other and fully
entangled with the environment. At the same time, the von Neumann
entropy of the radiation saturates.

If black holes $\mathcal B_L$ and $\mathcal B_R$ lack of d.o.f.,
they might be disentangled at the beginning of the evolution.
Therefore, the conditions leading to the negative density
difference $\tilde{S}_0$ cannot capture the dynamics of the
intermediate stage, and is not appropriate to the information
paradox for black holes -- see also Sec.~\ref{sec_pagecurve}.

Based on above observation, we argue that to obtain a Page curve
successfully in the context of eternal black holes, the essential
condition is to input sufficient d.o.f. into the black hole
subsystem, that will increase the entropy in (\ref{eq_QESinRad}).
In the weak tension limit as discussed in previous section, we
implement this by choosing the endpoint of HRT surface with large
$w_b$, which actually just transfer some d.o.f. from baths to the
black hole subsystem. Now once the backreaction is taken into
account, we have more strategies to enhance the d.o.f. on the
brane directly, which from our point of view should be more
natural to address the information paradox for the eternal black
hole. To realize it, the key point is to decrease the
$d$-dimensional Newton constant
\cite{Akal:2020wfl,Miao:2020oey,Chen:2020uac,Chen:2020hmv}.
Therefore, one possible way is to increase the tension on the
brane by adjusting the value of $\theta$ from $\pi/2$ to $0$ --
see (\ref{eq_a}) and (\ref{eq_a1}). Alternatively, one may
increase the effects of the intrinsic curvature term (DGP term) on
the brane by adjusting the DGP coupling $\lambda$ -- see
(\ref{eq_Area}).

In the remainder of this section, we will demonstrate that above
strategies works well in giving rise to a positive $\tilde{S}_0$,
even with $w_b=0$.

\subsection{The entropy density in the back-reacted spacetime} \label{sec_qei}

In this subsection, all free parameters will be fixed for
numerical analysis. To regularize the UV divergence of entropy,
we should introduce a UV cut-off. Since $\gamma_{pl}$ and
$\gamma_{tr}$ share the same asymptotic behavior, their difference
should be independent of the cut-off given that it is small enough. We find that the UV cut-off $y_\epsilon=1-1/100$ is good enough for numerical calculation.

From (\ref{eq_SR2}), the entropy density of the radiation subsystem $\tilde
S_{\mathcal{R}}$ is now determined by
\begin{align}
    \tilde S_{\mathcal{R}}=2\min \kc{\mathop{\text{ext}}\limits_{y_b}\left[\tilde S_{pl}(y_b) + \tilde S_{ _{DGP}}(y_b) \right], \tilde S_{tr}}.
\end{align}
Here $y_b$ is the intersection of $\gamma_{pl}$ with the
brane. Notice that $\gamma_{pl}$ is not a candidate of the QES
until taking the extremum.
$\mathop{\text{ext}}\limits_{y_b}\left[\tilde S_{pl}(y_b) + \tilde
S_{ _{DGP}}(y_b) \right]$ and $ \tilde S_{tr}$ are the entropy
density of two candidates respectively, while the entropy
density of the radiation subsystem is identified with the minimum of two
candidates. Next, we derive the expressions for the entropy
density of two candidates in a parallel way as presented in the
previous section.

For the trivial surface $\gamma_{tr}$ at $t=0$ as the QES, we introduce the parameterization $x=x(y)$, which leads to the corresponding entropy density
\begin{align}\label{eq_Str}
\tilde S_{tr}(0)=\frac{L^2}{4
G_N^{(4)}} \int_0^1 dy\frac{1}{\left(1-y^2\right)^2} \sqrt{Q_5
\left(\frac{4Q_2}{P(y)}+\frac{Q_4 \left(2 y (x(y)-1)^2
Q_3+x'(y)\right)^2}{(x(y)-1)^4}\right)}.
\end{align}
For numerical convenience, we have set ${L^2/(4 G_N^{(4)})}=1$
in the following discussion.

\begin{figure}
    \centering
    \subfigure[]{\label{fig_candidates}
        \includegraphics[width=195pt]{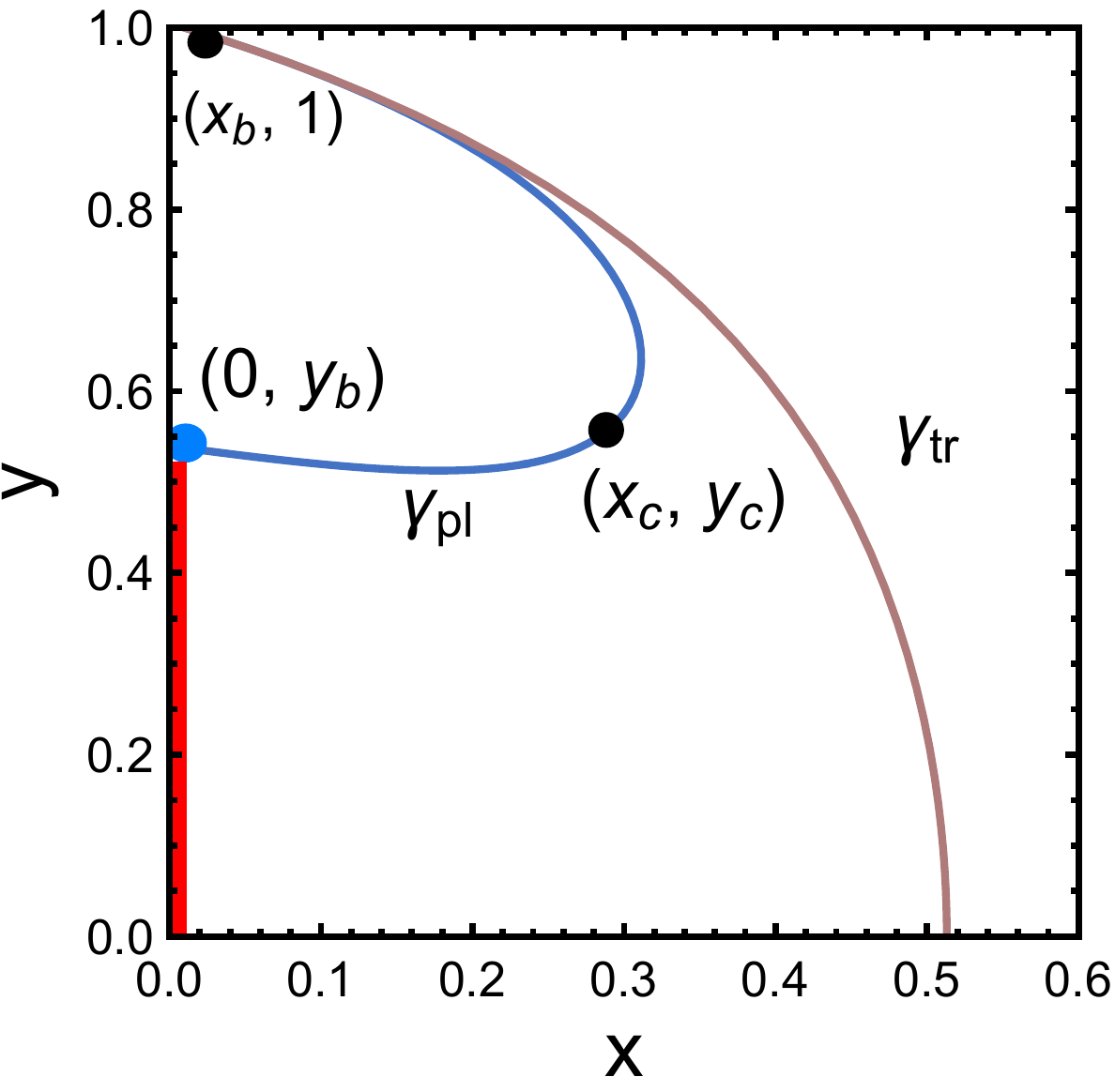}}
    \hspace{0pt}
    \subfigure[]{\label{fig_deltaSmu0}
        \includegraphics[width=215pt]{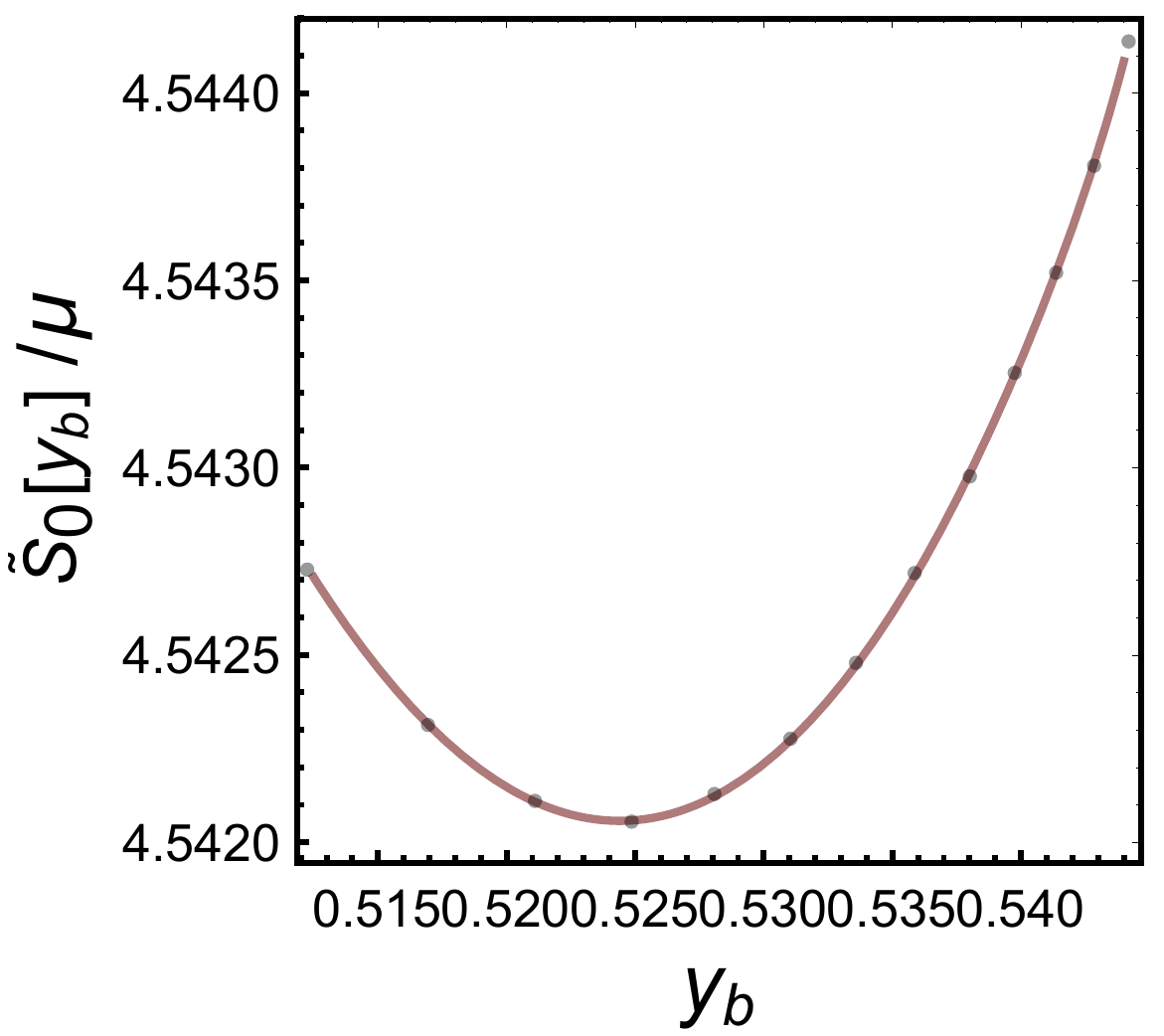}}
\caption{(a): For
$\{\theta,T_h/\mu,\lambda,x_b\}=\{\pi/4,0.19894,1,\frac{1}{100}\}$, the
surface penetrating the horizon  $\gamma_{tr}$ is colored in rose
gold, while the surface anchored on the brane $\gamma_{pl}$
and the corresponding contribution of the DGP term are colored in
blue. Both of them are anchored at $x=x_b$ on the conformal
boundary, while the half of island at $t=0$ is colored in red.
(b): For $\{\theta,T_h/\mu,\lambda,x_b\}=\{\pi/4,0.19894,1,\frac{1}{100}\}$,
the density difference between two surfaces is plotted
as a function of $y_b$, by which the entropy density difference $\tilde{S}_0/\mu$ can be figured out as the minimum, with
$\tilde{S}_0/\mu \approx 4.54206$.}
\end{figure}

As for the surface $\gamma_{pl}$ at $t=0$, we also introduce two
different parameterizations in different intervals. In $(x,y)$
plane, for the curve in $y \in [y_c,1]$ -- see
Fig.~\ref{fig_candidates}, we introduce $x=x(y)$ just as the
parameterization of the trivial surface $\gamma_{tr}$, while for
the curve in $x \in [x_c,1]$, we introduce $y=y(x)$ instead, with
$y'(x_c)=x'(y_c)^{-1}$. Finally, the density associated
with the surface $\gamma_{pl}$ anchored at $y_b=y(0)$ can be read
off from the integration procedure, with the following expressions
\begin{align}
 \tilde{S}_{pl}(y_b) =&\int_{y_c}^1 \frac{dy}{\left(1-y^2\right)^2} \sqrt{Q_5 \left(\frac{4Q_2}{P(y)}+\frac{Q_4 \left(2 y (x(y)-1)^2 Q_3+x'(y)\right)^2}{(x(y)-1)^4}\right)} +\int_{0}^{x_c} \frac{dx}{\left(y(x)^2-1\right)^2}
 \nonumber\\
  &\sqrt{Q_5 \left(\frac{4 Q_2 y'(x)^2}{P(y(x))}+Q_4 \left(4 Q_3^2 y(x)^2 y'(x)^2+\frac{4 Q_3 y(x) y'(x)}{(x-1)^2}+\frac{1}{(x-1)^4}\right)\right)},\label{eq_Spl}\\
  \tilde{S}_{ _{DGP}}(y_b)=&
  \frac{\lambda \sqrt{Q_5}}{1-y_{b}^2}\Big|_{x=0},\quad
  y_{b}=y(0).\label{eq_Area}
\end{align}
Notice that varying $(x_c, y_c)$ will subsequently change
$y_b$. Thus $\tilde{S}_{pl}(y_b)$ also depends on $y_b$.

The density difference between the surface
$\gamma_{tr}$ and the surface $\gamma_{pl}$ with the DGP
contribution is given by
\begin{equation}
  \tilde{S}_0[y_b]= 2\left(\tilde{S}_{pl}(y_b) + \tilde{S}_{DGP}(y_b) - \tilde{S}_{tr}(0)\right).
\end{equation}
Then the corresponding entropy density difference between two
candidates at $t=0$ is obtained by extremizing the above equation,
which is
\begin{align}\label{eq_deltaS}
  \tilde{S}_0 = \tilde{S}_0[y_{ _{QES}}] = \mathop{\text{ext}}\limits_{y_b}  \tilde{S}_0[y_b],
\end{align}
where $\tilde{S}_0[y_b]$ reaches it extremum at $y_b = y_{
_{QES}}$. The whole process is illustrated in
Fig.~\ref{fig_deltaSmu0}.

\begin{figure}
  \centering
  \subfigure[]{\label{fig_xbyb}
  \includegraphics[width=140pt]{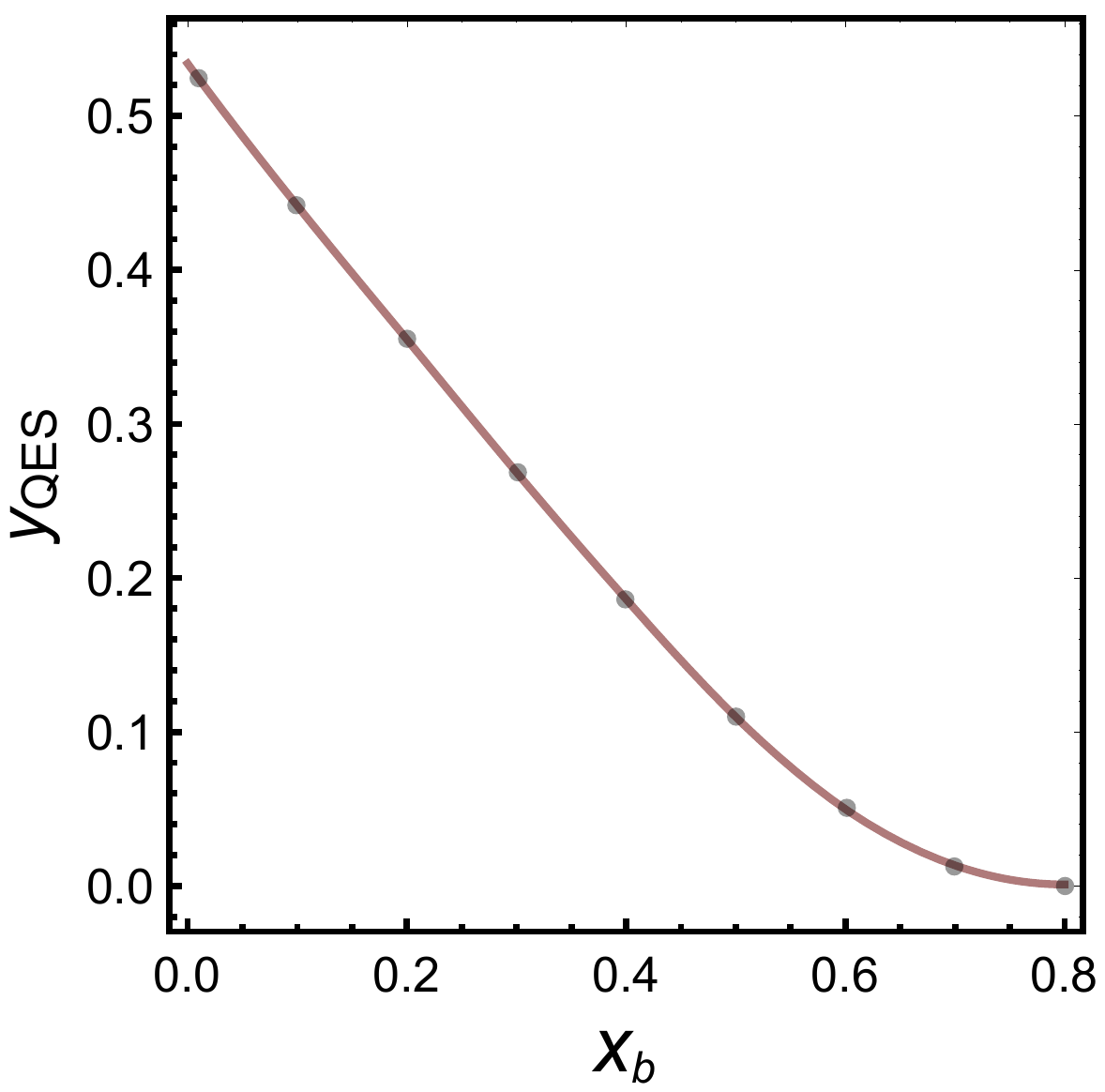}}
  \hspace{0pt}
  \subfigure[]{\label{fig_xbS}
  \includegraphics[width=142pt]{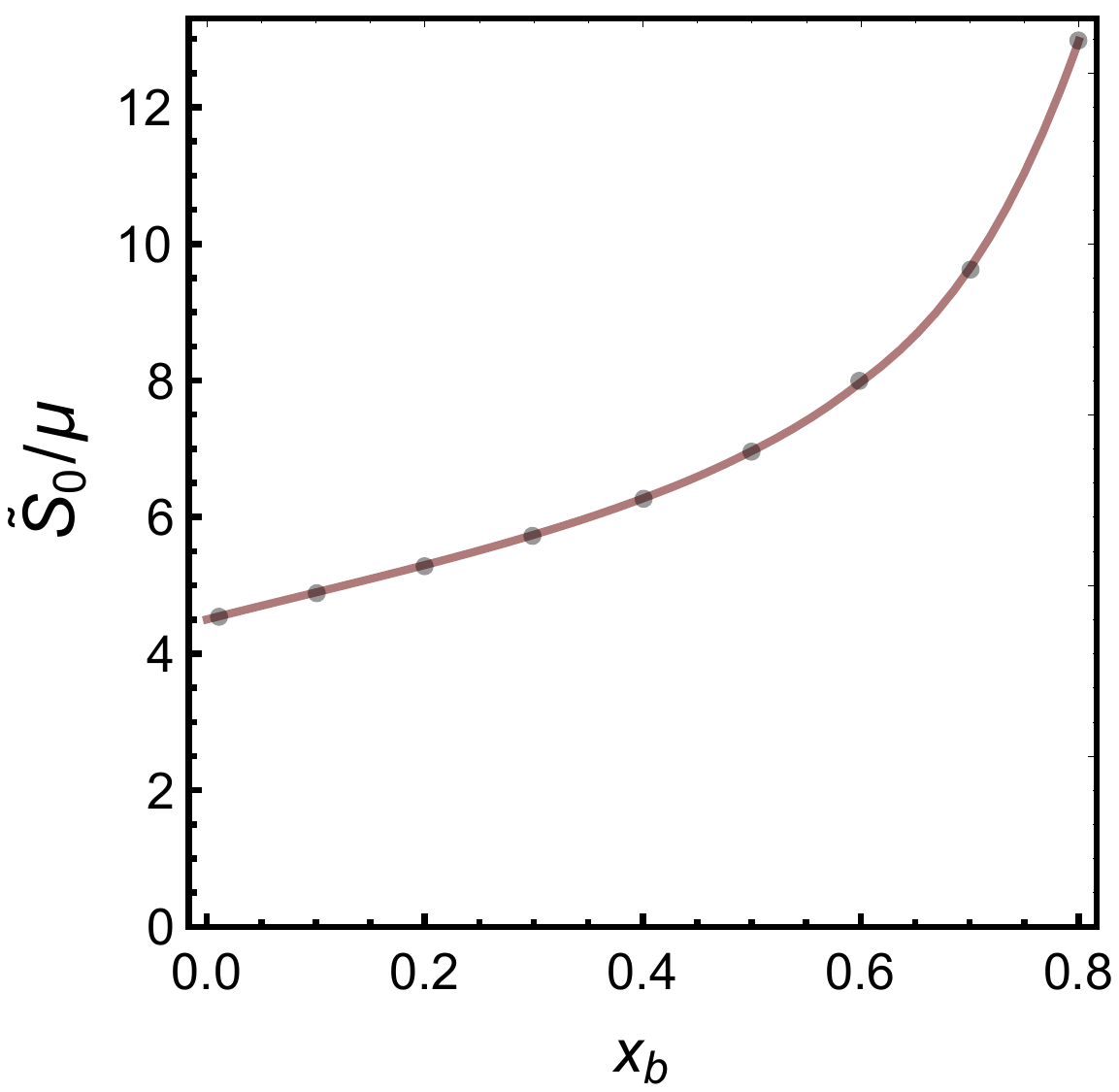}}
  \hspace{0pt}
  \subfigure[]{\label{fig_muyb}
  \includegraphics[width=136pt]{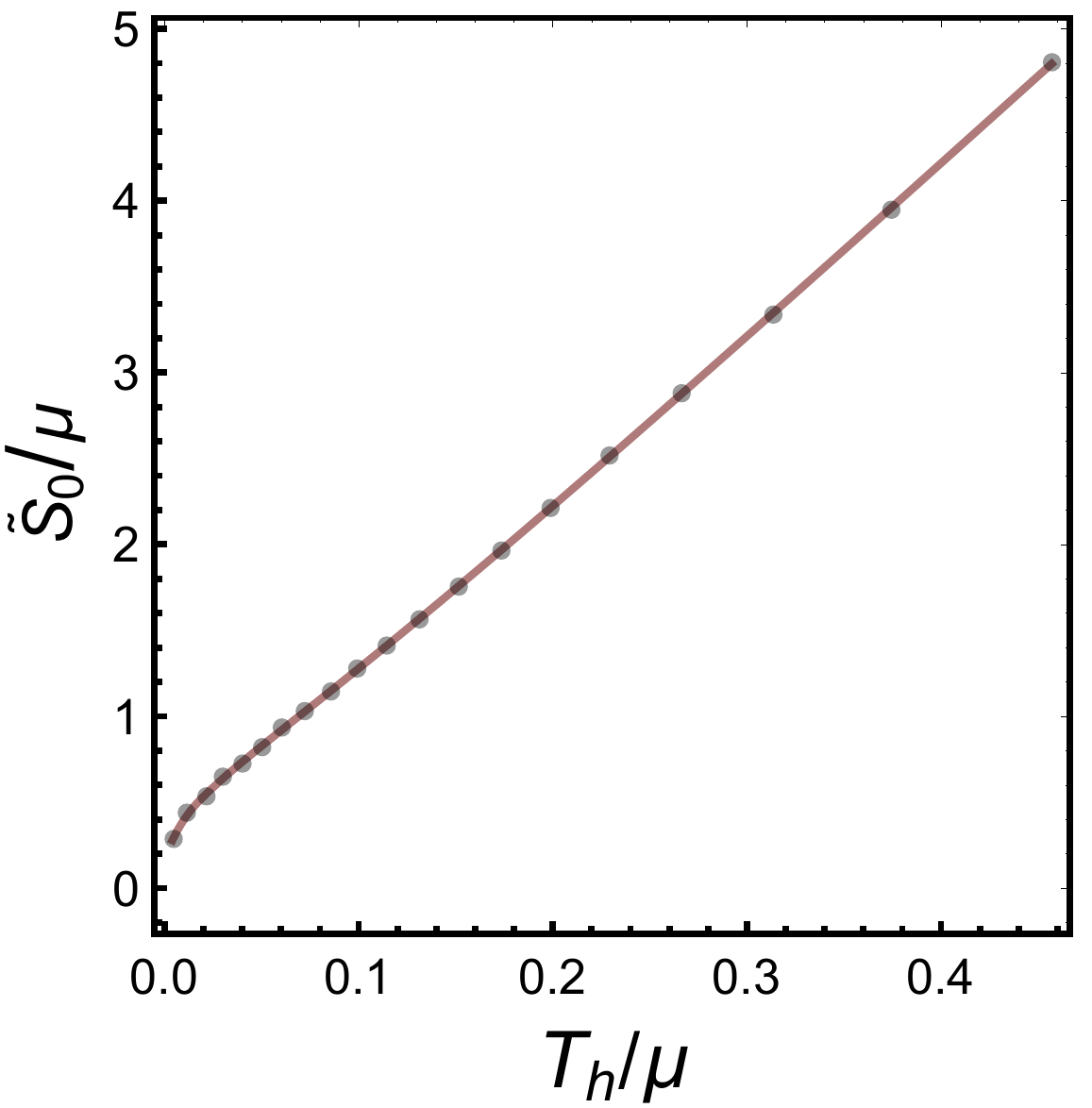}}\\
  \hspace{0pt}
  \subfigure[]{\label{fig_matter}
  \includegraphics[width=140pt]{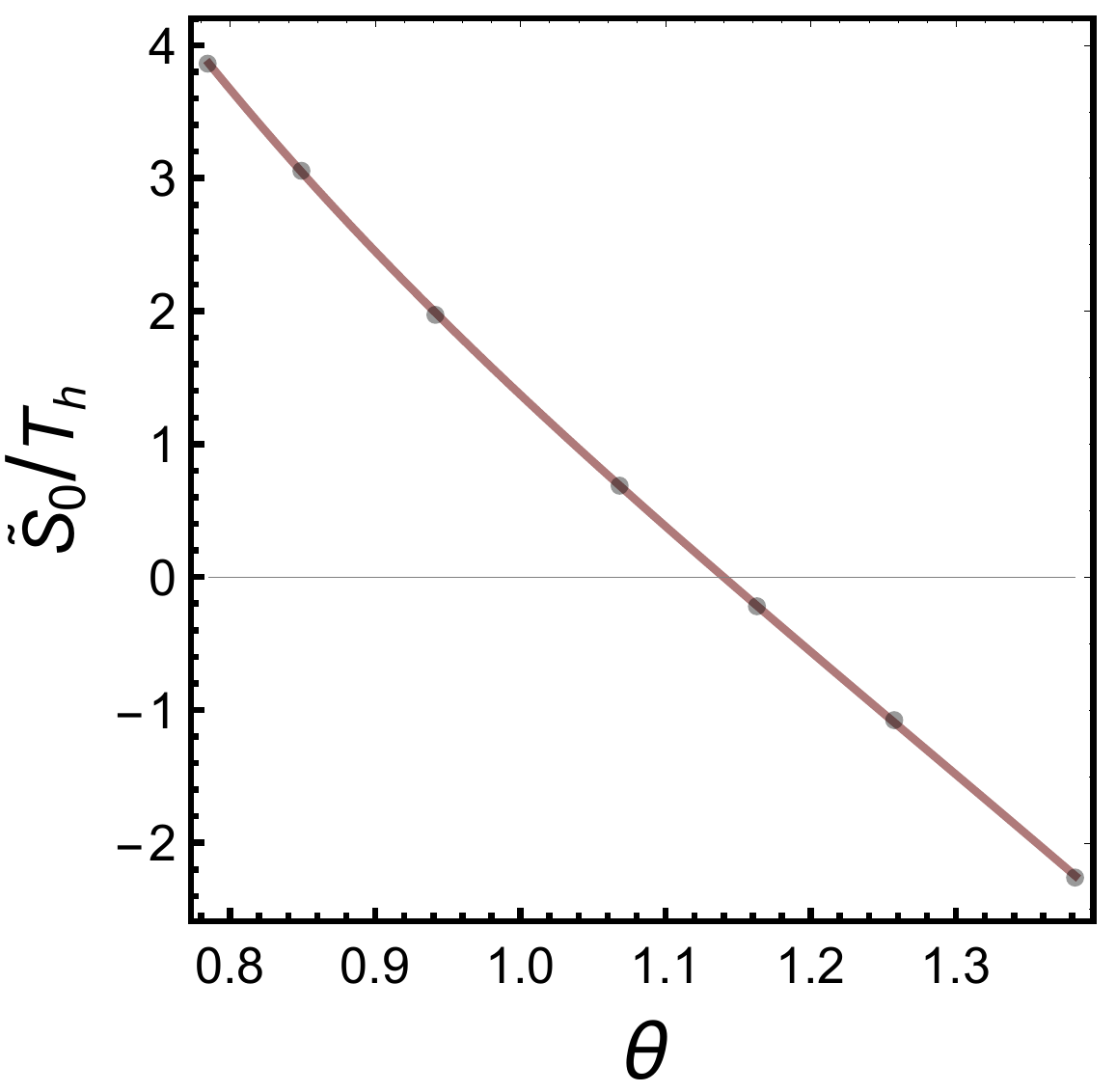}}
  \hspace{0pt}
  \subfigure[]{\label{fig_lambda}
  \includegraphics[width=142pt]{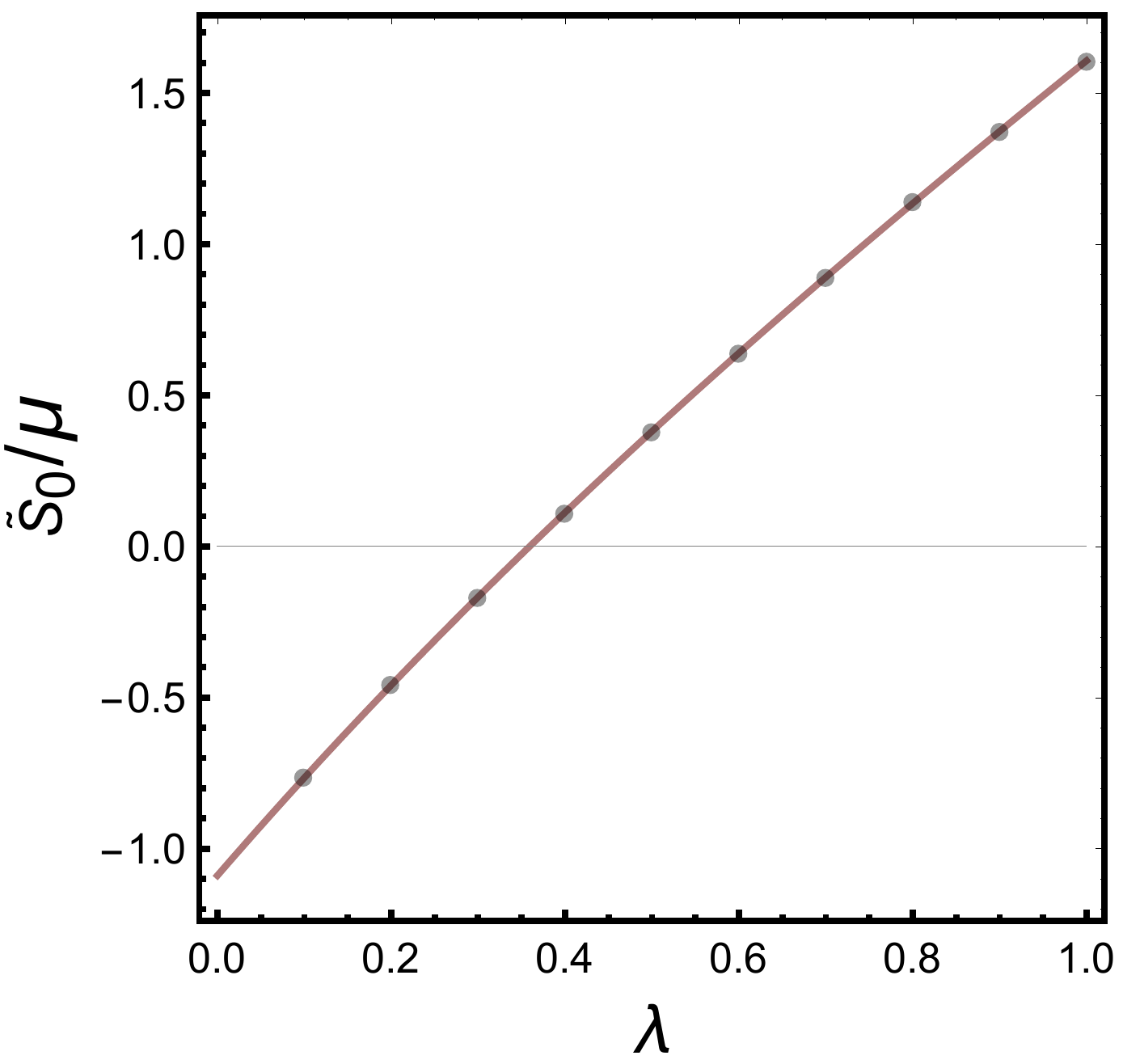}}
\caption{(a) The location of QES $y_{ _{QES}}$ as a function of
$x_b$, where $\{\theta,T_h/\mu,\lambda\}=\{\pi/4,0.19894,1\}$.
 (b) For $\{\theta,T_h/\mu,\lambda\}=\{\pi/4,0.19894,1\}$, the relation between the density difference
$\tilde{S}_0/\mu$ and the endpoint $x_b$ is plotted as the solid
curve. (c) For $x_b=1/100$, the relation between the density difference $\tilde{S}_0/\mu$ and $T_h/\mu$ is plotted as the solid curve. (d) For $\{d,\mu/T_h,\lambda,x_b\}=\{3,0,0,1/5\}$, the relation
between the density difference $\tilde{S}_0/T_h$ and $\theta$. (e)
For $\{d,\theta,T_h/\mu,x_b\}=\{3,\pi/2,0.19894,1/5\}$, the relation between
the density difference $\tilde{S}_0/\mu$ and the DGP coupling
$\lambda$.}\label{fig_Constraint}
\end{figure}

In Fig.~\ref{fig_deltaSmu0}, the difference $\tilde{S}_0/\mu$ is
positive at $y_{ _{QES}}\approx 0.524$. Therefore, the QES at
$t=0$ is the trivial surface $\gamma_{tr}$.
As Hawking radiation continues, the entanglement entropy keeps growing until
Page time $t_P$. Then, the surface $\gamma_{pl}$ dominates at
$t=t_P$ and the entropy saturates due to the phase transition.
After Page time, the island solution dominates and the island
$\mathcal{I}$ emerges in the entanglement wedge of the radiation subsystem.
Therefore, the d.o.f. on the island are encoded in the radiation subsystem by
quantum error correction process.

Similar to result as obtained in Sec.~\ref{sec_pagecurve}, the
island will stretch out of the horizon, as illustrated in
Fig.~\ref{fig_xbyb}. It indicates that the region beyond the
horizon will also be encoded in the radiation subsystem by the entanglement
wedge. Furthermore, the entropy density difference $\tilde S_0/\mu$ also
increases with the endpoint $x_b$, which is plotted in
Fig.~\ref{fig_xbS}.

Moreover, $\tilde{S}_0/\mu$ increases with Hawking temperature
$T_h/\mu$ as well, which manifests that the growth of
temperature allows more entanglement to be built up between the
black hole and radiation subsystem, as shown in
Fig.~\ref{fig_muyb}.

In comparison with picture in the weak tension limit, taking the
backreaction into account leads to the following significant
changes. The first is that for $x_b \rightarrow 0$, the location
of $y_{ _{QES}}$ approaches $0.52$ -- see Fig.~\ref{fig_xbyb},
which indicates that only the region near the horizon ($0<y<y_{
_{QES}}$) is likely to be encoded in the radiation subsystem. The second is that for $x_b
\rightarrow 0$, the density difference $\tilde{S}_0/\mu$ remains
positive, which clearly demonstrate that the Page curve can always be obtained, as long as the tension as well as the coupling is large enough.

Moreover, we remark that two strategies of increasing the difference of entropy density work independently, which are
illustrated in Fig.~\ref{fig_matter} and Fig.~\ref{fig_lambda}.
For the first, one may decrease $\theta$ from $\pi/2$ to $0$ but
fix $\lambda=0$,
then with more tension on the brane, the density difference will
turn to positive and the entropy starts to evolve.
Alternatively
one may increase the DGP coupling $\lambda$ but fix
$\theta=\pi/2$, the density difference will also turn
from negative to positive.

\section{Conclusions and Discussions}\label{sec_Conclusion}
In this paper, we have investigated the black hole information
paradox in the doubly-holographic setup for charged black holes in
general dimensions. In this setup, the holographic dual of a
two-sided black hole is in equilibrium with baths such that the
geometry keeps stationary during the evaporation.

In the weak tension limit, we have analytically calculated the entropy
of the radiation by figuring out the quantum extremal surface.
With appropriate choice of the endpoint of the HRT surface,
 the Page curve has successfully been obtained with the presence of an island on the
brane. The Page time is evaluated as well. For the neutral
background in general $(d+1)$ dimensions, the entropy grows
linearly at the rate proportional to $T_h^{d-1}$ due to the
evaporation and finally saturates at a constant which depends on
the size of the radiation system,
while for the near-extremal case in general $(d+1)$ dimensions, the entropy grows
linearly at the rate proportional to $T_h$. Specifically,
in $4D$ charged case, we have
plotted the Page curves under different temperatures. At high
temperatures, the black hole seems to ``evaporate'' more rapidly
than the cold one and will build more entanglement with the
radiation. While for the near-extremal black hole, the evolution
seems to be frozen, since the exchanging of Hawking modes takes
place extremely slowly.

It should be noticed that in the weak tension limit, to guarantee that the difference of entropy density $\tilde{S}_0$ is positive, one has to choose the
endpoint of the HRT surface away from the brane. As a matter of
fact, such a choice transfers d.o.f. from baths to the black hole
subsystem, leading to the fact that the entanglement between two
subsystems is mainly contributed from the CFT matter fields.

Keep going further, we have obtained the stationary solution with
backreaction by the standard DeTurck trick. Rather than
transferring d.o.f. from baths, we have proposed two strategies to
avoid the saturation of entropy at initial time. One is increasing
the tension on the brane and another is introducing an intrinsic
curvature term on the brane. Both increasing the DGP coupling $\lambda$ and decreasing the angle $\theta$ will relatively enlarge the central charge of the $(d-1)$-dimensional conformal defect, which is dual to the brane theory. Therefore, these two prescriptions are equivalent to enhancing the
d.o.f. on the brane directly and hence, are more appropriate to
address the information paradox for black holes.  As a result, the
Page curve can always be recovered, as long as the tension
$\alpha$ or the DGP coupling $\lambda$ is large enough.

Next it is very desirable to investigate the time evolution of the
entanglement entropy with backreaction, which involves in both
more d.o.f. on the brane and the dynamics of black holes, thus
beyond the DeTurck method. Furthermore, beyond the simple model
with doubly-holographic setup, the information paradox for
high-dimensional evaporating black holes is more complicated and
difficult to describe. Therefore, it is quite interesting to
develop new methods to explore the evaporation of black holes in
holographic approach.

\section*{Acknowledgments}
We are grateful to Shao-Kai Jian, Li Li, Chao Niu, Rong-Xing Miao, Cheng Peng, Shan-Ming Ruan, Yu Tian, Meng-He Wu, Xiaoning Wu, Cheng-Yong Zhang, Qing-Hua Zhu for helpful discussions. In particular, we would like to thank Hongbao Zhang for improving the manuscript during the revision,
and the anonymous referee for drawing our attention to the DGP
term on the brane. This work is supported in part by the Natural
Science Foundation of China under Grant No.~11875053, 12035016 and 12075298.
Z.~Y.~X. also acknowledges the support from the National
Postdoctoral Program for Innovative Talents BX20180318, funded by
China Postdoctoral Science Foundation.

\bibliographystyle{unsrt}

\bibliography{HwkRadi0222}

\begin{thebibliography}{10}

\bibitem{Page:1993wv}
Don~N. Page.
\newblock {Information in black hole radiation}.
\newblock {\em Phys. Rev. Lett.}, 71:3743--3746, 1993.

\bibitem{Page:2004xp}
Don~N. Page.
\newblock {Hawking radiation and black hole thermodynamics}.
\newblock {\em New J. Phys.}, 7:203, 2005.

\bibitem{Page:2013dx}
Don~N. Page.
\newblock {Time Dependence of Hawking Radiation Entropy}.
\newblock {\em JCAP}, 09:028, 2013.

\bibitem{hawking1974black}
Stephen~W Hawking.
\newblock Black hole explosions?
\newblock {\em Nature}, 248(5443):30--31, 1974.

\bibitem{Almheiri:2012rt}
Ahmed Almheiri, Donald Marolf, Joseph Polchinski, and James Sully.
\newblock {Black Holes: Complementarity or Firewalls?}
\newblock {\em JHEP}, 02:062, 2013.

\bibitem{Maldacena:2013xja}
Juan Maldacena and Leonard Susskind.
\newblock {Cool horizons for entangled black holes}.
\newblock {\em Fortsch. Phys.}, 61:781--811, 2013.

\bibitem{Almheiri:2018xdw}
Ahmed Almheiri.
\newblock {Holographic Quantum Error Correction and the Projected Black Hole
  Interior}.
\newblock 10 2018.

\bibitem{Penington:2019npb}
Geoffrey Penington.
\newblock {Entanglement Wedge Reconstruction and the Information Paradox}.
\newblock 5 2019.

\bibitem{Almheiri:2019psf}
Ahmed Almheiri, Netta Engelhardt, Donald Marolf, and Henry Maxfield.
\newblock {The entropy of bulk quantum fields and the entanglement wedge of an
  evaporating black hole}.
\newblock {\em JHEP}, 12:063, 2019.

\bibitem{Ryu:2006bv}
Shinsei Ryu and Tadashi Takayanagi.
\newblock {Holographic derivation of entanglement entropy from AdS/CFT}.
\newblock {\em Phys. Rev. Lett.}, 96:181602, 2006.

\bibitem{Lewkowycz:2013nqa}
Aitor Lewkowycz and Juan Maldacena.
\newblock {Generalized gravitational entropy}.
\newblock {\em JHEP}, 08:090, 2013.

\bibitem{Engelhardt:2014gca}
Netta Engelhardt and Aron~C. Wall.
\newblock {Quantum Extremal Surfaces: Holographic Entanglement Entropy beyond
  the Classical Regime}.
\newblock {\em JHEP}, 01:073, 2015.

\bibitem{Almheiri:2019hni}
Ahmed Almheiri, Raghu Mahajan, Juan Maldacena, and Ying Zhao.
\newblock {The Page curve of Hawking radiation from semiclassical geometry}.
\newblock {\em JHEP}, 03:149, 2020.

\bibitem{Chen:2019uhq}
Hong~Zhe Chen, Zachary Fisher, Juan Hernandez, Robert~C. Myers, and Shan-Ming
  Ruan.
\newblock {Information Flow in Black Hole Evaporation}.
\newblock {\em JHEP}, 03:152, 2020.

\bibitem{Chen:2019iro}
Yiming Chen.
\newblock {Pulling Out the Island with Modular Flow}.
\newblock {\em JHEP}, 03:033, 2020.

\bibitem{Balasubramanian:2020hfs}
Vijay Balasubramanian, Arjun Kar, Onkar Parrikar, G\'abor S\'arosi, and
  Tomonori Ugajin.
\newblock {Geometric secret sharing in a model of Hawking radiation}.
\newblock 3 2020.

\bibitem{Hashimoto:2020cas}
Koji Hashimoto, Norihiro Iizuka, and Yoshinori Matsuo.
\newblock {Islands in Schwarzschild black holes}.
\newblock {\em JHEP}, 06:085, 2020.

\bibitem{Krishnan:2020fer}
Chethan Krishnan.
\newblock {Critical Islands}.
\newblock 7 2020.

\bibitem{Almheiri:2020cfm}
Ahmed Almheiri, Thomas Hartman, Juan Maldacena, Edgar Shaghoulian, and
  Amirhossein Tajdini.
\newblock {The entropy of Hawking radiation}.
\newblock 6 2020.

\bibitem{Alishahiha:2020qza}
Mohsen Alishahiha, Amin Faraji~Astaneh, and Ali Naseh.
\newblock {Island in the Presence of Higher Derivative Terms}.
\newblock 5 2020.

\bibitem{Gautason:2020tmk}
Fridrik~Freyr Gautason, Lukas Schneiderbauer, Watse Sybesma, and L'arus
  Thorlacius.
\newblock {Page Curve for an Evaporating Black Hole}.
\newblock {\em JHEP}, 05:091, 2020.

\bibitem{Almheiri:2019yqk}
Ahmed Almheiri, Raghu Mahajan, and Juan Maldacena.
\newblock {Islands outside the horizon}.
\newblock 10 2019.

\bibitem{Penington:2019kki}
Geoff Penington, Stephen~H. Shenker, Douglas Stanford, and Zhenbin Yang.
\newblock {Replica wormholes and the black hole interior}.
\newblock 11 2019.

\bibitem{Almheiri:2019qdq}
Ahmed Almheiri, Thomas Hartman, Juan Maldacena, Edgar Shaghoulian, and
  Amirhossein Tajdini.
\newblock {Replica Wormholes and the Entropy of Hawking Radiation}.
\newblock {\em JHEP}, 05:013, 2020.

\bibitem{Almheiri:2019psy}
Ahmed Almheiri, Raghu Mahajan, and Jorge~E. Santos.
\newblock {Entanglement islands in higher dimensions}.
\newblock {\em SciPost Phys.}, 9(1):001, 2020.

\bibitem{Geng:2020qvw}
Hao Geng and Andreas Karch.
\newblock {Massive Islands}.
\newblock 6 2020.

\bibitem{Chen:2020uac}
Hong~Zhe Chen, Robert~C. Myers, Dominik Neuenfeld, Ignacio~A. Reyes, and Joshua
  Sandor.
\newblock {Quantum Extremal Islands Made Easy, Part I: Entanglement on the
  Brane}.
\newblock {\em JHEP}, 10:166, 2020.

\bibitem{Chen:2020hmv}
Hong~Zhe Chen, Robert~C. Myers, Dominik Neuenfeld, Ignacio~A. Reyes, and Joshua
  Sandor.
\newblock {Quantum Extremal Islands Made Easy, Part II: Black Holes on the
  Brane}.
\newblock {\em JHEP}, 12:025, 2020.

\bibitem{Hernandez:2020nem}
Juan Hernandez, Robert~C. Myers, and Shan-Ming Ruan.
\newblock {Quantum Extremal Islands Made Easy, PartIII: Complexity on the
  Brane}.
\newblock 10 2020.

\bibitem{Akal:2020wfl}
Ibrahim Akal, Yuya Kusuki, Tadashi Takayanagi, and Zixia Wei.
\newblock {Codimension two holography for wedges}.
\newblock 7 2020.

\bibitem{Miao:2020oey}
Rong-Xin Miao.
\newblock {An Exact Construction of Codimension two Holography}.
\newblock 9 2020.

\bibitem{Takayanagi:2011zk}
Tadashi Takayanagi.
\newblock {Holographic Dual of BCFT}.
\newblock {\em Phys. Rev. Lett.}, 107:101602, 2011.

\bibitem{Chu:2018ntx}
Chong-Sun Chu and Rong-Xin Miao.
\newblock {Anomalous Transport in Holographic Boundary Conformal Field
  Theories}.
\newblock {\em JHEP}, 07:005, 2018.

\bibitem{Miao:2018qkc}
Rong-Xin Miao.
\newblock {Holographic BCFT with Dirichlet Boundary Condition}.
\newblock {\em JHEP}, 02:025, 2019.

\bibitem{Headrick:2009pv}
Matthew Headrick, Sam Kitchen, and Toby Wiseman.
\newblock {A New approach to static numerical relativity, and its application
  to Kaluza-Klein black holes}.
\newblock {\em Class. Quant. Grav.}, 27:035002, 2010.

\bibitem{Dias:2015nua}
Óscar~J.C. Dias, Jorge~E. Santos, and Benson Way.
\newblock {Numerical Methods for Finding Stationary Gravitational Solutions}.
\newblock {\em Class. Quant. Grav.}, 33(13):133001, 2016.

\bibitem{Geng:2020fxl}
Hao Geng, Andreas Karch, Carlos Perez-Pardavila, Suvrat Raju, Lisa Randall,
  Marcos Riojas, and Sanjit Shashi.
\newblock {Information Transfer with a Gravitating Bath}.
\newblock 12 2020.

\bibitem{Randall:1999vf}
Lisa Randall and Raman Sundrum.
\newblock {An Alternative to compactification}.
\newblock {\em Phys. Rev. Lett.}, 83:4690--4693, 1999.

\bibitem{Dvali:2000hr}
G.R. Dvali, Gregory Gabadadze, and Massimo Porrati.
\newblock {4-D gravity on a brane in 5-D Minkowski space}.
\newblock {\em Phys. Lett. B}, 485:208--214, 2000.

\bibitem{Karch:2000ct}
Andreas Karch and Lisa Randall.
\newblock {Locally localized gravity}.
\newblock {\em JHEP}, 05:008, 2001.

\bibitem{Nozaki:2012qd}
Masahiro Nozaki, Tadashi Takayanagi, and Tomonori Ugajin.
\newblock {Central Charges for BCFTs and Holography}.
\newblock {\em JHEP}, 06:066, 2012.

\bibitem{Hartman:2013qma}
Thomas Hartman and Juan Maldacena.
\newblock {Time Evolution of Entanglement Entropy from Black Hole Interiors}.
\newblock {\em JHEP}, 05:014, 2013.

\bibitem{Carmi:2017jqz}
Dean Carmi, Shira Chapman, Hugo Marrochio, Robert~C. Myers, and Sotaro
  Sugishita.
\newblock {On the Time Dependence of Holographic Complexity}.
\newblock {\em JHEP}, 11:188, 2017.

\end{thebibliography}

\end{document}